Saarland University

Center for Bioinformatics

Master's Program in Bioinformatics

**Master's Thesis**

# A Pre-Docking Filter Based on Image Recognition

submitted by

**Eva Kiszka**

on September 30, 2011

Supervisor

PD Dr. Michael C. Hutter

Advisor

PD Dr. Michael C. Hutter

Reviewers

PD Dr. Michael C. Hutter
Prof. Dr. Volkhard Helms

# Statement

I hereby confirm that this thesis is my own work and that I have documented all sources used.

Saarbrücken, September 30, 2011

Eva Kiszka

# Acknowledgements

First and foremost I am indebted to PD Dr. Michael C. Hutter. He provided me with a relevant and innovative subject for this thesis. Always being available when needed, he also was a great advisor. I highly recommend him to anyone seeking for a committed supervisor.
Sincere thanks are given to Prof. Dr. Volkhard Helms for inspiring topical discussions and advice. Likewise, I would like to thank him for agreeing to review my thesis.

I would like to express my gratitude to the open source developers of the software used in the context of writing this thesis (e.g. AutoDock, Eigen, and Collabtive).

Furthermore I thank Karin Jostock from the Center for Bioinformatics' secretariat and examination office for her amazingly professional help in all administrative issues.

Last but not least I would like to state that the support, patience, and encouragement given by my husband Philipp Kiszka and my family were a crucial factor for the successful completion of my master's degree. Thank you all.

# Abstract

Molecular docking is a central method in the computer-based screening of compound libraries as a part of the rational approach to drug design. Although the method has proved its competence in predicting binding modes correctly, its inherent complexity puts high demands on computational resources. Moreover the chemical space to be screened is prohibitively large. Therefore the application of filtering prior to docking is a promising concept.

We implemented a pre-docking filter based on the tangent distance algorithm originally conceived for optical character recognition. The challenging transfer of the method from two-dimensional to three-dimensional data was achieved by representing the molecular structure by a set of density maps extracted from different views of the compound.

Additionally, our program applies a binary classification using principal component analysis. Ligand and binding pocket are aligned according to their centroidal axes, enabling a size-based filtering for the purpose of enriching the dataset regarding ligands before docking.

The evaluation of our program via redocking produced RMSD values between 8Å and 25Å, indicating that the tangent distance approach is not suited for optimizing the orientation of a ligand and binding pocket. Investigating probable explanations lead to the conclusion that a likely cause for these results is the method's known inability to approximate large transformations.

A validation of the principal component analysis alone performed better: Tests on a dataset of 170 ligands and 6,435 decoys yielded a sensitivity of 0.81, while keeping the runtime within a reasonable timeframe (1 to 4 seconds). The dataset's enrichment increased from 2.64% to 2.82%.

# Table of Contents

# 1 Introduction

## 1.1 About this Thesis

This master's thesis is the final step within a graduation project for the Master of Science (M. Sc.) degree program in Bioinformatics at Saarland University, Saarbrücken, Germany. The project is carried out within the Chair of Computational Biology held by Prof. Dr. Volkhard Helms. His group, including PD Dr. Michael C. Hutter, studies the structure, dynamics, energetics, interactions, and mechanisms of biomolecules and subcellular systems. Research topics span fields as diverse as computational chemistry, theoretical biophysics, and biotechnology.

The chair is part of the Center for Bioinformatics (CBI), a highly interdisciplinary scientific institution, in which the Faculty of Medicine and the Faculty of Natural Sciences & Technology I & III of Saarland University cooperate with the Max Planck Institute for Computer Science, Saarbrücken, the Fraunhofer Institute for Biomedical Engineering IBMT, St. Ingbert, and the German Research Center for Artificial Intelligence DFKI, Saarbrücken.

Well in accordance with the CBI's mission statement *From diseases to therapies with bioinformatics*, the aim of this thesis was the development and evaluation of a pre-docking filter – a bioinformatics tool with the potential to speed up the process of drug discovery. The CBI's vision of interdisciplinary research is furthermore reflected in the approach of transfering an algorithm best known from optical character recognition (OCR) to molecular modeling.

## 1.2 Context

The process of drug discovery has changed drastically over the last decades [K09, pp. 9-10].

In the past researchers relied on trial-and-error-based methods in order to identify chemical substances with a desired clinical effect. All tests had to be carried out manually either in vitro or in vivo. Nowadays it is possible to apply a more systematic approach to the task of drug discovery: the so-called rational drug discovery, which includes process automation and beyond that in most cases also in silico, i.e. computer-based methods.

The first step of this process is the identification of a biological target with a therapeutic value, e.g. a protein with a known influence on a disease-related process. On top of that, a molecule which is suited as a target must be *druggable*, which means it is able to bind to a small molecule, which subsequently changes the target's activity. Another term for target is *receptor*, which refers to its role as the *receiving* part in the formation of a complex with a smaller molecule.

The second step of rational drug discovery is to identify such small molecules, which are also called ligands (ligare [Latin], to bind) because of their ability to bind the target. If the three-dimensional (3D) structure of the target is known, it is possible to tackle this task in silico. The computer-based search in vast libraries of chemical structures (like e.g. the ZINC database [I05]) for ligands is called virtual screening (VS).

The main advantage of VS over screening assays is its speed. Thus, VS enables researchers to scan in reasonable time large parts of compound databases for molecules that can be synthesized or purchased, and tested. However, it is a difficult task to computationally assess the probability of a small molecule to bind to a given target.

There exist two different basic approaches to VS:

1. Ligand-based methods
2. Structure-based methods

Although we decided to separately describe each of the following approaches to VS, it is common to combine them in more complex implementations as e.g. done for docking and MD simulations by Hritz et al. 2008 [H08].

## 1.2.1  Ligand-Based Virtual Screening

The methods of ligand-based VS require the availability of a set of ligands with different structures, which are known to bind to a given receptor ([H07]). Thus no structural information about the target is needed. The analysis of chemical and structural similarities of these ligands allows for identifying traits of the receptor, which might be important for its binding mode. From this information, it is possible to deduce a pharmacophore model of the receptor (see also Lemmen & Lengauer 1999 [L99]), which is an abstract description of its binding pocket including spatial and electro-static features.

Common approaches to ligand-based VS according to Klebe 2000 [K00] include:

- Feature trees
- Molecular fingerprints
- 3D descriptors

Since our approach does not have a connection to ligand-based methods, we will not further elaborate on those.

## 1.2.2   Structure-Based Virtual Screening

If the 3D structure of the target is known from nuclear magnetic resonance (NMR) spectroscopy, X-ray crystallography, or homology modeling, then usually structure-based VS is the preferred method.

The knowledge of the spatial conformation and chemical composition of a receptor allows for computational prediction of potential ligands and binding modes. This involves docking – a *prediction method for the orientation of one molecule to a second when bound to each other to form a stable complex*, according to an appropriate and relatively short definition given by Lengauer & Rarey 1996 [L96].

## 1.3   Docking

While it is imaginable to perform docking by computer-based molecular dynamics (MD) simulations, this is uncommon because of the enormous computational cost involved. However, researchers have experimented with it, because firstly ligand flexibility is inherent in this approach and secondly its modeling of the process of molecular recognition and induced fit is assumed to be very close to how the molecules interact in the real world (see also Nola et al. 1994 [N94], Nakajima et al. 1997 [N97], Mangoni et al. 1999 [M99]).

Another possible way of predicting binding modes during structure-based VS is by using so-called shape complementarity or shape-matching methods. The common idea of these approaches is to focus on geometrical features of the given molecules' surfaces and try to find an optimal docked pose in terms of fitting these surfaces together.

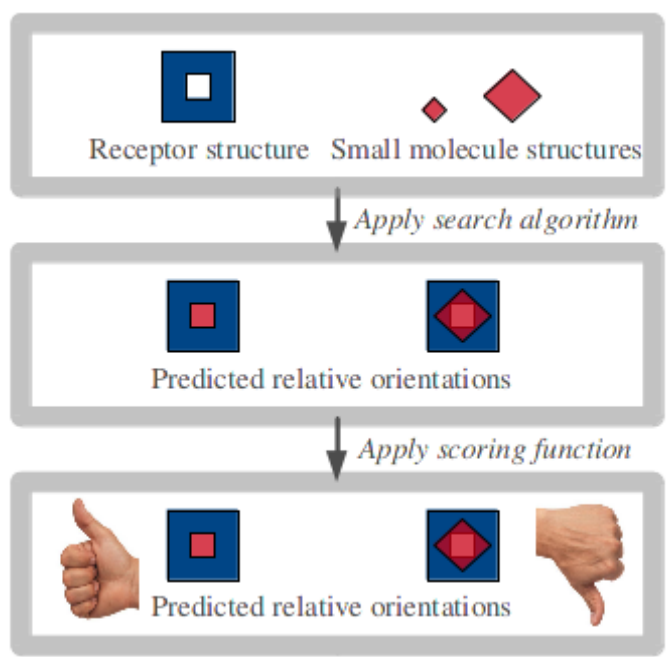

*Figure 1: General flow of the docking procedure. The receptor (blue) has a binding site (white hole) in which the docking software tries to fit the given small molecules (red) during the search step. Applying a scoring function to the resulting relative orientations yields a specific value making the poses comparable.*

A high-level outline of shape complementarity approaches to docking includes the following steps (see also Figure 1):

1. Retrieve 3D structure for receptor and potential ligand(s)
2. Apply search algorithm to sample space of possible orientations
3. Apply scoring function to assign ranks to possible orientations

Shape complementarity approaches will henceforth be referred to simply as *docking* since this is what is usually meant when this term is used. A more detailed introduction to docking-related searching, scoring, and ranking will be given later in this thesis, using as an example one of the docking tools we are going to present in the next chapter.

The benefits of docking however, are not restricted to VS, but the method is also applicable to

- X-ray crystallography,
- lead optimization,
- combinatorial library design, and
- chemical mechanism studies.

## 1.3.1  Available Software

### 1.3.1.1  Overview

While different docking tools implement different searching and scoring algorithms, there are yet more distinguishing features which have to be taken into account when deciding which one to use. The following list is not complete, but it gives a feeling for the complexity of the task of chosing the best option:

- **Availability / cost:**  open source software  ↔  *in-between*  ↔  commercial software
- **Focus of docking:**  protein-ligand  ↔  protein-protein  ↔  combinatorial
- **Flexibility / rigidity:**  rigid ligand or receptor  ↔  flexible ligand or receptor

Although there are more than two dozen tools available for protein-ligand docking according to Moitessier et al. 2008 [M08a], the market for such software has consolidated over the years and now is clearly dominated by less than 10 computer programs (see Table 1).

The following in-depth discussion of the elements of molecular docking will mainly focus on their implementation in AutoDock (AD), since this software package was used in the evaluation of our work for this thesis. Before we delve into the specifics of searching, scoring, and ranking, we will therefore give an overview of some basic facts worth knowing about AD.

### 1.3.1.2  AutoDock

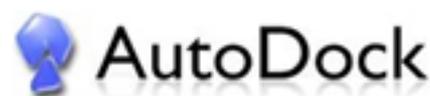

*Figure 2: AD logo.*

AD's maintainers The Scripps Research Institute and Olson Laboratory made the molecular modeling package open source and hence freely available for any kind of use in 2009. Since then, it seems to have become one of the most widespread, if not even the single most widespread docking software in academia (see [B11], [P08], [R07]), even though this is difficult to quantify. The tool was successfully applied in the discovery of the first clinically approved human immunodeficiency virus (HIV) 1 integrase inhibitor [S04a].

The package consists of three main components:

1. AutoGrid: Pre-calculate a set of grids describing the protein
2. AutoDock: Dock the ligand to the set of grids calculated by AutoGrid
3. AutoDockTools (ADT)

ADT is a graphical user interface for the package. In this way the developers offer a convenient method to prepare the coordinate and other files needed in a specific format as input for AD and to analyze the results of a docking run. But it is also possible to use AD without ADT, according to Morris et al. 2008 [M08b], by interacting with the program via the command line.

| Tool | Maintainer | Availability* | Flexibility | | Search | Scoring |
| | | | Protein | Ligand | | |
|------|-----------|---------------|-----------|--------|--------|---------|
| AutoDock | The Scripps Research Institute | Free | Single amino acids | Stochastic | SA**, GA*** | FF****, Empirical, Knowledge-based |
| DOCK | UCSF | Free | - | Systematic | IE***** | FF |
| FlexX | BioSolveIT | Commercial | Ensemble of protein structures | Systematic | IE | Empirical, Knowledge-based |
| GOLD | CCDC | Commercial | Single amino acids | Stochastic | GA | Empirical, Knowledge-based |
| Glide | Schrodinger | Commercial | - | Systematic | SA, IE | Empirical, Knowledge-based |
| ICM | MolSoft | Commercial | Side chains | Stochastic, Deterministic | SA | FF |
| QXP/Flo+ | n/a | Free | - | Stochastic | SA | FF |
| Surflex-Dock | Tripos | Commercial | - | Systematic | IE | Empirical, Knowledge-based |

*Table 1: An overview of the most widely used protein-ligand docking software. * for commercial use only; ** SA: Simulated annealing. *** GA: Genetic algorithm. **** FF: Force field; ***** IE: Incremental extension. Information taken from [C10], [M06], [R96], [V03], [H04], [A94], [M97], [J03]; supplemental data from Lazarova [L08].*

## 1.3.2  Search Algorithms

The major challenge during the development of a docking search algorithm is finding a way to deal with or bypass the exhaustive exploration of the search space. According to Morris et al. 2010 [M10], *The ideal procedure would find the global minimum in the interaction energy between the substrate and the target protein, exploring all available degrees of freedom (DOF) for the system.*

For orienting two molecules to each other in every possible way at a given level of granularity is a task with six DOF's: three for translation in x, y, and z direction, and another three for rotation around the x, y, and z axis. Most docking software additionally takes ligand flexibility into account, which again adds to the computational complexity. Here it becomes clear why so few programs offer a solution accounting for different protein conformations (see also [C07]).

Established ways of dealing with the problem of the enormous search space are applying simulated annealing (SA), incremental extension (IE), or genetic algorithms (GA). AutoDock 3 and later versions offer searching via an optimized SA or two different implementations of GA's according to Morris et al. 1998 [M98]. All established methods rely on a discretization of the search space: The software discretizes the available binding pocket on a lattice and only considers discrete conformational changes in terms of flexibility.

### 1.3.2.1  Simulated Annealing

Finding an optimal pose of the ligand in the binding site of the receptor requires evaluation of the free energy function of interaction with respect to the parameters resulting from the given pose, and a comparison of different poses with regard to their energetics. Optimizing the poses is a problem that can be solved by SA, a probabilistic metaheuristic. The algorithm does not necessarily find the global minimum regarding the interaction energy of available poses, it does however find a good approximation in relatively short time.

The algorithm performs local optimization of a given solution until convergence by making slight changes to the pose, which lower the interaction energy. In order to avoid getting stuck in a local minimum, it performs larger random jumps from time to time (see Figure 2).

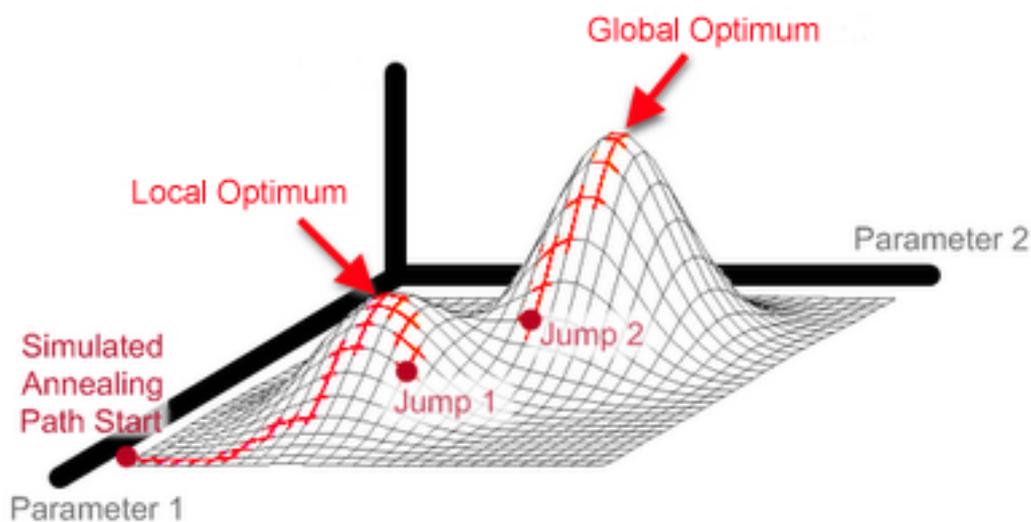

*Figure 3: Illustration of the idea behind chaotic jumps in SA. After reaching a local optimum, a metaheuristic is applied: It induces jump 1, which still does not escape the local optimum – but jump 2 does and finds the global optimum. Modified from Dama 2008 [D08], by permission from Max Dama.*

#### 1.3.2.2   Genetic Algorithms

These algorithms are inspired by natural evolutionary processes such as mutation. Just as SA, they apply random elements in order to solve a global optimization problem by approximation. Over the years, their role in docking has become more important, because they proved to be especially apt when applied to combinatorial explosion problems ([M98]).

The current state of a pose is defined by a translation, orientation, and conformation of the given ligand with respect to the corresponding receptor. By transferring the terms of crossover, mutation, selection, and generations from natural evolution to an implemented *evolution* of ligand translation, rotation, and conformation, it is possible to create new variants with better *fitness.* Fitness in this context means a lower interaction energy.

### 1.3.3   Scoring and Ranking

A *good* candidate ligand does not only fulfill the requirement of fitting into the binding site. If the aim is to develop a new drug, it should also be *drug-like* – a concept including properties such as bioavailability and metabolic stability [V08]. Last but not least it is necessary that the compound be a strong selective binder of the receptor, which means it interacts specifically with this protein and less with other targets, and it binds the receptor tightly, which is referred to as high affinity [G02].

Estimating the likelihood of a small molecule being a good binder for a given target is the role of scoring in the context of docking. The core of most scoring functions is an approximation of the pose's energetical favorability. For this purpose they implement molecular mechanics force fields (FF's), e.g. the AMBER FF used in AutoDock 4.2 (see Morris et al. 2009 [M09b]). The FF evaluation is sometimes combined with other descriptors, like knowledge-based methods deriving the statistical potentials for interactions from known complexes [Z08].

Just like the task of searching described in chapter 1.3.2, the task of discriminating the quality of poses is challenging, too. Estimating intramolecular energetics in a way that comes close enough to the real non-covalent binding energy while at the same time being feasible with the given hardware and time is difficult: According to Morris et al. 2010 [M10] one of the basics of FF's is the evaluation of all of the system's atoms' pair-wise interaction energies. AD includes terms for dispersion/repulsion, hydrogen bonding, electrostatics, and desolvation in this evaluation. Then again, on a higher level, these terms are not only evaluated once per atom, but for the bound and the unbound case respectively. Additionally, the FF considers a term for the conformational entropy lost upon binding.

Such energy-based scoring functions allow for the approximation of the binding affinity as defined by the change in the system's Gibbs free energy $\Delta G$:

$$\Delta G = -RT \; lnK$$

*Equation 1: G = Gibbs free energy, R = universal gas constant, T = temperature, K = equilibrium constant.*

When all visited poses have been assigned a scoring by the scoring function, it is possible to rank the poses according to their scores. This is the last step of a docking run, whereupon the user finally is presented an ordered list of poses along with the corresponding scores.

## 1.3.4  Relevance

While the development of computational methods for molecular modeling brought great improvements to many areas of life sciences, the results of VS have yet to live up to the initial promise of largely replacing empirical screening (see also Shoichet 2004 [S04b]).
Structure-based methods shone in some cases (e.g. Villoutreix et al. 2009 [V09]), and VS is an established complement to empirical screening – but still not a substitute. Some point out that docking can only achieve good results, if the user has sufficient expert knowledge to choose suitable parameters [C09], others even contemplate that VS might have reached its peak [S10]. However, aside from the fact that maximally improving the established methods makes sense anyway, there are two important reasons why we think that the outlook for VS and docking methods in particular remains positive:

1. Moore's Law (Moore 1965 [M65]) says that the density of transistors on integrated circuits doubles approx. every two years. His prediction proved to be true for several decades already – with an end not yet in sight [D10]. Moreover the research field of high performance computing has made available additional resources e.g. by transfering parts of the computational burden to the graphics processing unit (GPU) (Sukhwani 2005 [S05b], Larsson et al. 2011 [L11]).

    This trend constantly widens the computational constraints, which limit the accuracy of in silico modeling and simulation. While the calculating capacity grows, it is possible to run algorithms that are ever closer to the course of events in the real world (i.e. for the case of bioinformatics the course of events in vivo) within reasonable periods of time. It enables accounting for more parameters reflecting the real state of an object of interest. Examples are outlined in Vendruscolo & Dobson 2011 [V11] and Mitchell & Matsumoto 2011 [M11].

2. Figure 4 shows that the amount of structural data available as an input for screening in drug design is ever growing. Although this growth has been slowing down slightly from its initial exponential increase according to Levitt 2006 [L06], it still makes digitalization more and more essential. While integration and consolidation of the new protein information represent just two more challenges for bioinformaticians (see also Andreeva et al. 2008 [A08a]), on the other hand the added data also opens up new perspectives. Having more and more high-resolution protein structures at hand will also result in a growing interest in improved docking methods.

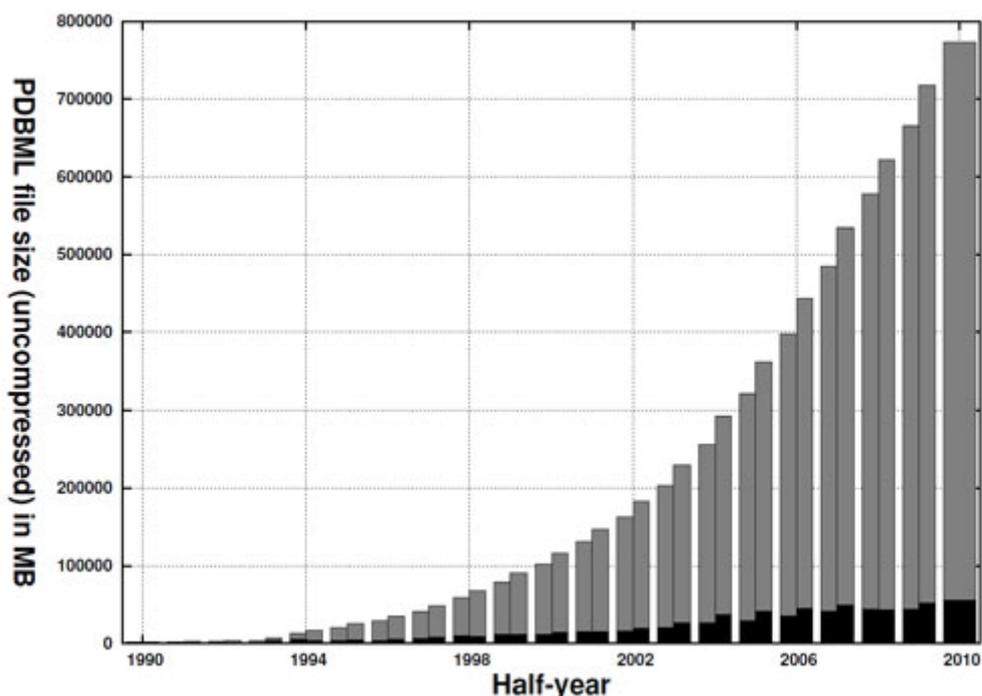

*Figure 4: Growth of the Protein Data Bank (PDB) over the past two decades. Grey bars: Size of all structures deposited. Black bars: Size of the structures added in the respective year. From Anders & Nicola 2011 [A11].*

## 1.4   Basics of Optical Character Recognition

Since our approach implements an algorithm well-established in OCR, we will now give a short introduction to this field.

The objective of OCR is the automated and mostly computer-aided recognition of printed or handwritten text. The resulting output typically is a text file in a format like e.g. ASCII or UTF-8. Combining several fields of information technology, OCR mainly depends on computer vision and pattern recognition [I91]. Since we experimented with a pattern recognition algorithm, we will only elaborate on this component of the OCR process.

The abstract purpose of pattern recognition can be described as the classification of objects into categories. In the context of OCR, these objects are single characters or digits, and the classification algorithm is supposed to map each object onto one of a given set of classes, representing e.g. the digit *9* or the character *b*. Basically, OCR is about shape-matching algorithms.

A practical field of application of OCR-related pattern recognition is handwritten digit recognition for the interpretation of zip codes on postal envelopes in preparation of automatic mail sorting.

Many groups developing approaches to the problem used as input for training and validation a dataset provided by the United States Postal Service (USPS) (see e.g. [K02], [K00a]). This *USPS handwritten digit recognition task* contains normalized, segmented 16x16 pixels (i.e. a total of 256 pixels) grayscale images of handwritten digits (0 to 9), joined together in a training set of 7,291 images and a second set of 2,007 images for testing purposes [K04]. On the level of data structures, one such image is usually represented as a 16x16 matrix, and it can be interpreted as a column vector: The vector is the result of *unrolling* the corresponding matrix by taking column after column starting from the left side, and attaching it to the lower end of the vector. Hence, a *m* x *n* matrix will result in a column vector with $m \cdot n$ rows or dimensions, wherein each dimension is a feature describing the brightness of a specific pixel [M09a]. So, a vector resulting from unrolling a 16x16 matrix will have 256 dimensions.

A classification requires the availability of a set of classes. In the case of handwritten digit recognition this set equals $C = \{0, 1, ..., 8, 9\}$. The digit to be classified is compared to a set of handwritten digits (usually containing several slightly varied instances of each of the classes), whose classification is known. These representative objects are called *prototypes* (see Figure 5).

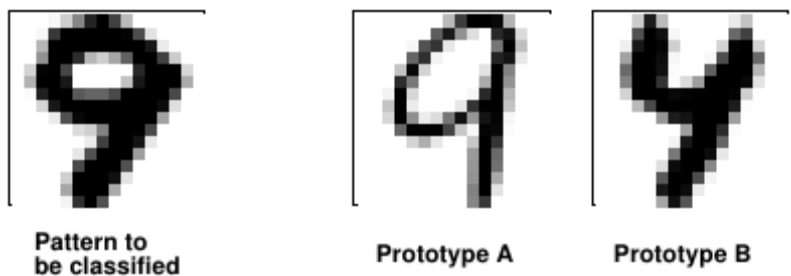

*Figure 5: Starting point of handwritten digit classification. The pattern to be classified (left) is a 16x16 pixels grayscale image obviously representing a 9. This is not so obvious for a technical device. Two prototypes representing the classes 9 (Prototype A, middle) and 4 (Prototype B, right) are given as examples. From Simard et al. 1998 [S98].*

When it comes to classifying a pattern according to the classes in *C*, a central step is the composition of a matrix *D* for each of the classes. The matrix is supposed to represent parts of the allowed variety of the corresponding class. For this purpose one takes all $N_V$ vectors *V* representing the prototypes of a class $i \in C$ and lines them up as the columns of a new matrix $D_i$ (example: see Figure 6), where $D_i \in \mathbb{R}^{(256 \times N_V)}$.

The next step is finding out which class a given pattern *d* should be assigned to, i.e. to which class it has the smallest distance. Using mathematical terms, we would like to know whether d is a linear combination of the columns of $D_i$, or if it is not, then how close it is to being one. The pattern will be assigned to the class of whose columns it is a linear combination or where it comes closest to being a linear combination of the columns. [M09a]

$$D_5 = \begin{bmatrix} | & | & | & \dots & | \\ 5 & 5 & 5 & \dots & 5 \\ | & | & | & \dots & | \end{bmatrix}$$

*Figure 6: Matrix representing the class 5. The columns are made up of column vectors each of which represents one known prototype of the class. From Mazack 2009 [M09a].*

Finding the answer to this means solving a least-squares problem [Z99], which is a crucial step in handwritten digit recognition. Successful approaches to nearest-neighbor classification problems like the one described above usually include the application of a distance measure or *metric*. We will explain two such approaches in the following subchapters.

## 1.4.1 Naive Approach: Euclidean Distance

The most direct way of approaching the task at hand would be to measure the distance with the Euclidean metric. Applying the Euclidean distance to an elementary case where one wants to determine the distance between two points, means calculating the length of the straight line segment between them. The Euclidean distance $D_{Euclid}$ between two vectors *p* and *q* of length *n* is defined as follows [D09]:

$$D_{Euclid}(p,q) = \|p-q\| = \sqrt{(p_1-q_1)^2 + (p_2-q_2)^2 + \dots + (p_n-q_n)^2}$$

*Equation 2: Euclidean distance between two vectors.*

An important property of metrics is their behaviour with regard to invariants. An invariant is a trait of an object that does not change upon certain transformations. An example: When a *9* is rotated by 10°, it should still be recognized as a member of the class *9*. However, when the same symbol is rotated by 180°, we expect another result: the assignment to the class *6*. This exemplifies, that the allowed transformations are limited to a certain range.

Typical transformations in handwritten text are horizontal translation (shift), vertical translation, rotation, shear, scale, and line thinning. Therefore an appropriate metric for handwritten digit recognition should be invariant with respect to all of the above-mentioned deformations. [W05] However, the Euclidean metric is not: Even a shift by three pixels leads to an enormous growth of the square root of the sum of squares of pixel-to-pixel difference, as illustrated in Figure 7. This flaw would lead to large classification errors, hence this metric is not applicable to OCR [D00].

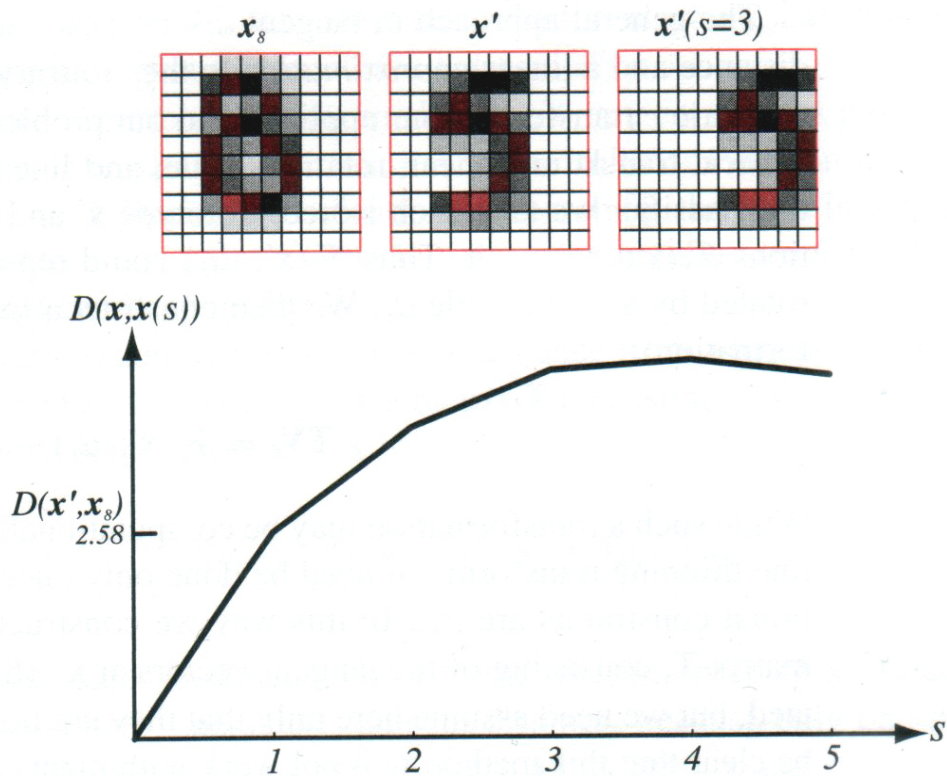

*Figure 7: Curve progression of the Euclidean metric illustrating its dependence on shift. The pattern to be classified is x' – a 5. The available prototypes are an 8 ($x_8$) and another 5, shifted by three pixels to the right (x'(s=3)). While it is obvious that the pattern should be classified as being closer to its shifted copy, the Euclidean distance (denoted by D) would say otherwise, because it is sensitive to horizontal translation: D(x', $x_8$) = 2.58 while D(x', x'(s=3)) ≈ 5.0. The curve shows how the Euclidean distance (y-axis) changes upon shifting the pattern by s pixels (x-axis). From Duda et al. 2000 [D00].*

## 1.4.2  Advanced Approach: Tangent Distance

In 1993, Simard et al. introduced a novel method to pattern recognition: *tangent distance* (TD) [S93]. Just as the Euclidean metric, TD is a distance measure that can be applied to points in Euclidean *n*-space, thus being utilizable to compute the distance between two 256-dimensional vectors, for example. In contrast to the Euclidean metric, TD is invariant to reasonably small transformations. It has already proved high performance in OCR [V98] and the classification of diverse other data, such as medical X-ray photographs ([K03], [L04]), video streams of American sign language [Z05], images of the German finger-spelling alphabet [D06], and speech [M01].

The following subsumptive containment hierarchy, deduced from Duda et al. 2000 [D00], puts TD within a larger context:
TD ⊂ Nearest-neighbor methods ⊂ Non-parametric methods ⊂ Classification ⊂ Pattern recognition

In order to solve the problem of handwritten digit recognition, a metric is needed, that is not just invariant to several transformations, but even to combinations of those: A digit may for example be rotated by 10° *and* shifted by 3 pixels to the right, but even then the optimal classifier is expected to put it into the same category as before the changes.

Simply trying to undo the combined translations is not an option, since they already can be computationally intensive on their own: For example the rotation of a digit and interpolation to a new grid is $O(n^2)$. Additionally, the parameters of the transformations (e.g. the angle of rotation or the pixels of translation) are not known in advance. [D00]

An abstract way to capture the behaviour of a vector undergoing multiple deformations is to say that the represented object describes manifolds in a high-dimensional space. Finding the shortest distance between a given unrolled image matrix and the usually non-linear space describing the allowed transformations is very difficult for two reasons (see also Simard et al. 1998 [S98]):

1. An algorithm will encounter many local minima regarding the distance between the curved surface and the point representing the pattern to be classified.
2. The computational burden is prohibitive more often than not because the procedure has to be reiterated for each prototype.

The TD approach avoids these pitfalls: It is applicable as a linear approximation to the curved surface, thus solving the problem of local minima. And then it combines arbitrary transforms in one representation encompassing the different prototypes of a given class.

While the Euclidean metric defines the distance between two points, TD can be referred to as the shortest distance between two planes. These planes are spanned by one of the integral parts computed by the TD algorithm: the *tangent vectors* (TV's). This is how they are constructed according to [D00]:

We assume that our problem allows for a total of *i* different transformations $t_i(x'; α_i)$, where *x'* is the prototype to which the transform is applied and $α_i$ is the respective parameter. In order to construct the classifier, we apply each of the *i* transformations to each of the $N_v$ prototypes. The subsequent construction of the TV's implements this equation:

$$TV_i = t_i(x'; α_i) - x'$$

*Equation 3: Calculation of a tangent vector TV.*

Afterwards, all the TV's resulting from different transformations of a prototype x' will be assembled into a matrix *T*. The matrix then contains linear approximations to all possible transformations. The following figures illustrate the interrelation between TD and true transformations (Fig. 8), and what happens, when several TV's are applied to a prototype at the same time (Fig. 9).

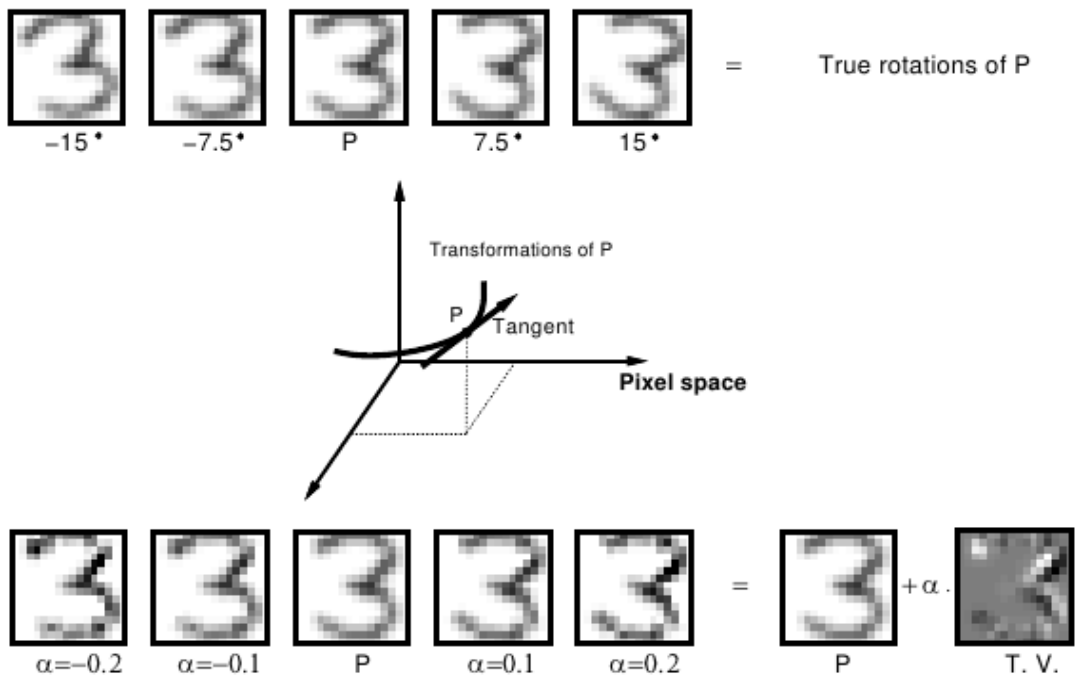

*Figure 8: Rotation of a prototype* P *and approximation by a TV representing rotation. The upper line of images shows the original* P *(middle) and its true rotations by steps with a difference of 7.5° in between each. The image line on the bottom shows the outcome of a linear combination of* P *and a TV accounting for rotation multiplied by parameter* α. *The results are similar to the upper line, which makes clear that TV's are a way to generate approximations of sufficient quality. The illustration in the middle clarifies the conceptual connection between the real transformations (curved line /* Transformations of P*) and the approximations resulting from TV application (straight arrow /* Tangent*). Both meet in the point of the untransformed prototype or the prototype plus a linear combination of the TV with parameter 0. From Simard et al. 1993 [S93].*

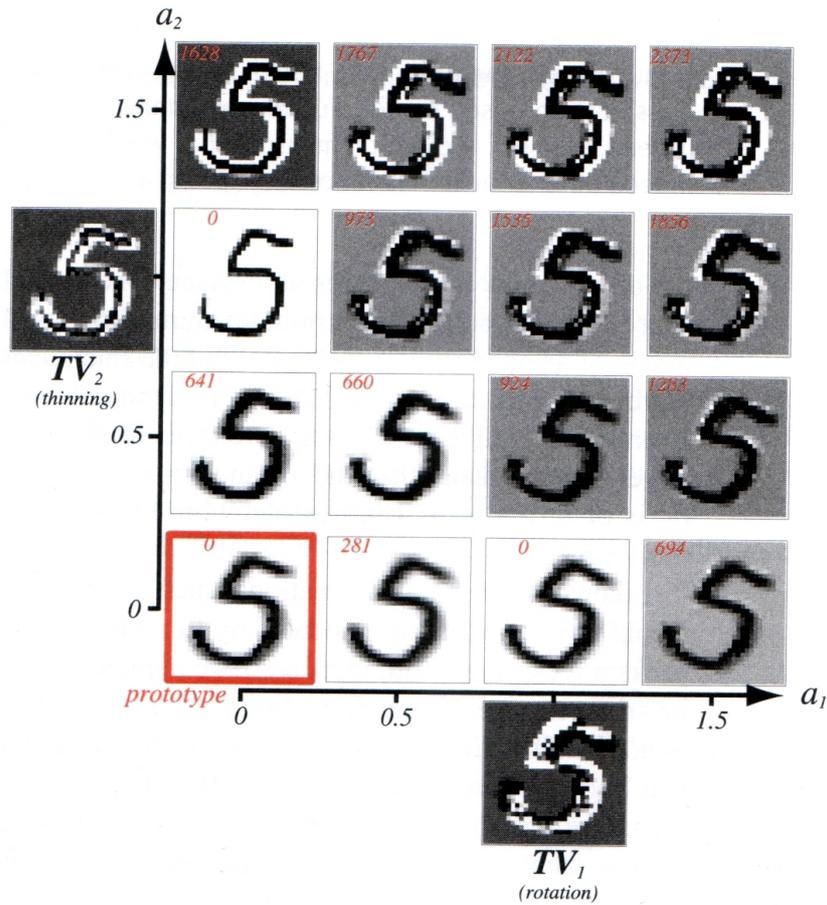

*Figure 9: Matrix of variations of a prototypic 5 resulting from adding linear combinations of two TV's to the prototype. The original prototype is played in the lower left square (red border). The TV's are located outside of the coordinate system: $TV_2$ is the outcome of applying a line thinning algorithm to the prototype and then subtracting the original prototype from its transformed version. $TV_1$ is a TV accounting for rotation by a small angle. Each pattern in the matrix represents the prototype plus a linear combination of the two TV's with parameters $\alpha_1$ (x-axis) for $TV_1$ and $\alpha_2$ (y-axis) for $TV_2$. From Duda et al. 2000 [D00].*

As soon as the matrix *T* of *i* TV's has been constructed, it is possible to compute the TD between a given test point *x* and our *x'*. This equation calculates the TD as the Euclidean distance between *x* and the tangent space of *x'*:

$$D_{Tangent}(x',x) = \min_{a}\left[\|(x' + Ta) - x\|\right]$$

*Equation 4: Calculation of the tangent distance TD.*

The only issue remaining is to find values for the parameter vector *a*, that minimize the TD. This linear least-squares problem can be solved by simply discretizing the search space and iterating over it – or more elegantly even by iterative gradient descent or matrix methods [D00].

# 2 Idea

Limitations in terms of available time and other resources still prevent researchers from simply screening all compounds available in virtual libraries – even if they have state-of-the-art computers and sophisticated molecular modeling software at hand to support them. Accelerating the process of docking would at least alleviate this problem.

One possible acceleration is an additional stage of filtering prior to docking, provided that the time saved during docking is longer than the time consumed by the application of the filter itself. Several methods of pre-docking filtering are quite common, for example quantitative structure-activity relationship (QSAR) modeling [A08b], pharmacophore screening [C11], or tests for drug-likeness like Lipinski's rule of five [T09].
Furthermore it could make sense to optimize a ligand's conformation and orientation before handing it over to the docking software.

The aim of this thesis was the development of a shape-matching-based pre-docking filter based on these two ideas:

- **1st step: Filtering**
  Avoid forwarding ligands to the docking tool, that are much too large for the available binding pocket in the first place
- **2nd step: Optimizing**
  Use tangent distance minimization as a measure of vague shape-matching to rotate and translate the ligand, so it overlaps with the binding pocket to the greatest possible extent

We will explain these ideas and the individual steps to be taken in the following subchapters.

## 2.1 Evaluation of the Binding Pocket

Before starting filtering and optimizing, a certain amount of preparatory work is required. Since we took an approach, in which the available binding pocket plays a central role, we first thing we needed to do is evaluate the shape of this free space within the binding site of the given receptor.

The user defines the outer boundaries of the considered area by specifying a cuboid by

- the coordinates of its center,
- the number of grid points to add from the center in each dimension, and
- the spacing between the points.

In the next step, we create a *negative imprint of the binding pocket* (in the following referred to as the binding pocket for brevity) within this cuboid: We fill up the free space with spheres. Some of it is occupied by the receptor's atoms. Taking these atoms' place requirements into consideration leads to the insight, that it is not sufficient to just not put spheres onto the grid points for whose exact coordinates a protein atom is present. It is necessary to account for the protein's and spheres' van der Waals (vdW) surface.

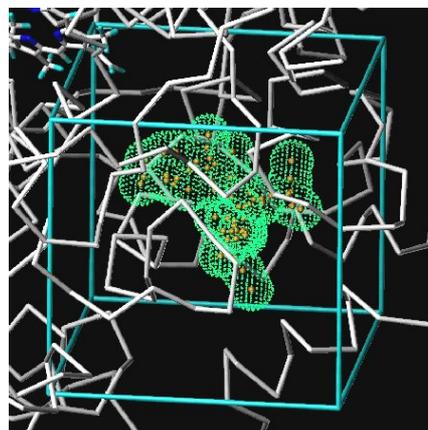

*Figure 10: Grid box filled up with spheres. Grey: Wireframe model of the receptor. Turquoise: Borders of the grid box. Green: Surface of the spheres. From Mozziconacci [M02].*

## 2.2   Alignment and Filtering

One of the ideas behind our filter is the application of principal component analysis (PCA), a classical statistical procedure (see e.g. [S05a]), to our setting. PCA is an appropiate means to find patterns in high-dimensional data – such as e.g. atomic coordinates – and was invented by Pearson in 1901 [P01].

A PCA reveals the principal components or *centroidal axes* of a set of data. Their number is equal to or smaller than the number of dimensions of the data. For a molecule for example, there will be 3 or less principal components, since it has atomic coordinates with a x-, y-, and z-dimension.

All centroidal axes are lines going through the data's center of mass or *centroid*, which is located at the point defined by averaging over each dimension's data. The first principal component minimizes the overall squared distance to all data points. One can say that this *line is as close to all of the data as possible* [S03]. In the case of a molecule, this line shows the direction of the longest stretch of the atoms. All following centroidal axes are orthogonal to each other. They represent the remaining less striking patterns in the data.

Figure 11 illustrates the concepts by giving an example based on two-dimensional (2D) data.

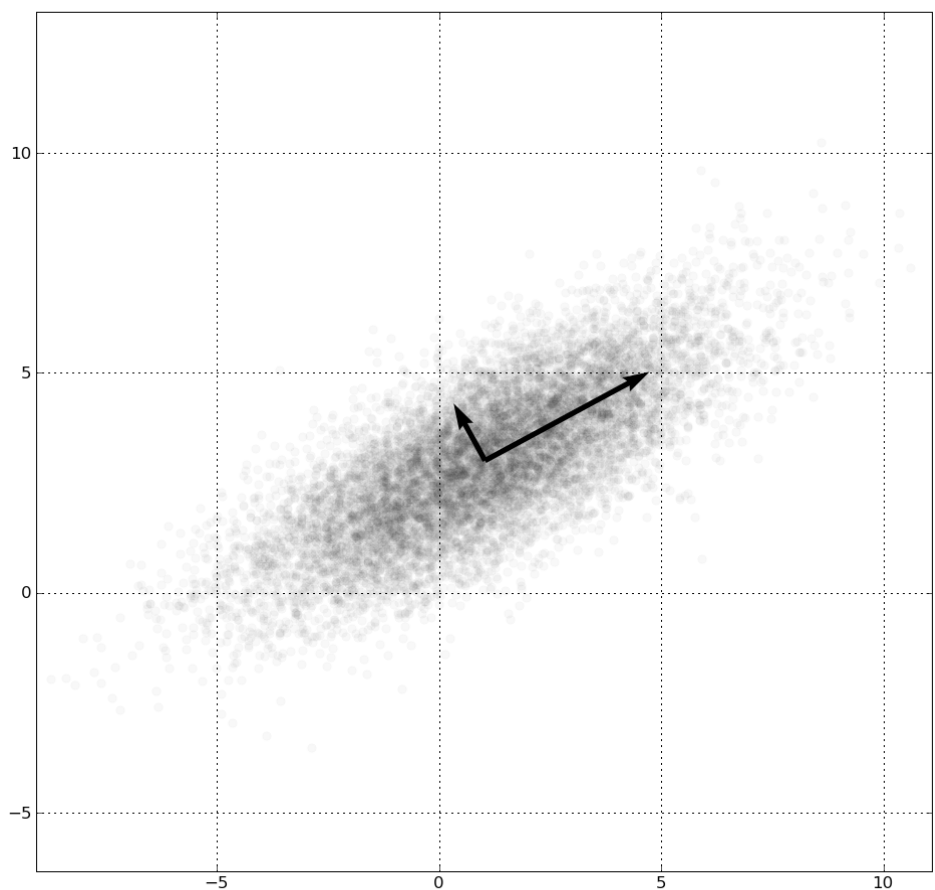

*Figure 11: PCA of a multivariate Gaussian distribution with a standard deviation of 3 in the direction of the longer arrow and of 1 in the orthogonal direction of the shorter one. The arrows correspond to the eigenvectors of the 2x2 covariance matrix scaled by the squareroot of the respective eigenvalue and shifted to the center of mass of the distribution. From Wikimedia Commons [W09].*

Applying PCA to a ligand and binding pocket includes overlapping their centroids and allows for aligning them along their centroidal axes. The resulting new orientation entails two opportunities, which we intended to exploit:

1. Comparing the ligand's coordinates furthest away from the centroid with the outermost atoms of the binding pocket might by a way to filter out ligands, that are too large for the given binding pocket.

2. There is a chance that the equally centered and axes-aligned ligand and binding pocket are a better starting point for docking than their initial random orientation.

## 2.3   Optimization of the Ligand Orientation

There is an interesting analogy between the problem sets of OCR and docking: Both include a number of allowed transformations. The classification of a symbol in OCR should be invariant to shift, shear, scale, and so on. The docking of a potential ligand starting from different initial orientations should produce the same result throughout. This behaviour is due to the problem-inherent invariants with respect to the following transformations and combinations of those:

- Translation in x-direction
- Translation in y-direction
- Translation in z-direction
- Rotation around x-axis
- Rotation around y-axis
- Rotation around z-axis

Our idea was to take advantage of this parallel and try to apply TD to optimize the ligand's orientation towards the binding pocket prior to docking.

There are discrepancies between the two settings calling for further elaboration:

While OCR solves a problem in 2-space, we have to deal with three dimensions. Our approach to solving this conflict was to create 2D representations of our objects of interest. We created density maps of the ligand and binding pocket by counting the number of atoms within each row of grid points. This procedure was reiterated from six different views: from front, left, above, back, right, and below. Then we computed the TD's for all allowed combinations of those and transformed the ligand according to the view that yielded the lowest TD.

In OCR, the prototypes are characters. TD is used as a measure to compare these characters to other, already classified characters. We wanted to compare ligands to binding pockets, not ligands to other ligands. Nevertheless it could make sense to approach the task in this way, since a minimization of the TD between a ligand and the binding pocket should still lead to an improved orientation: It overlaps the places where the ligand takes up most space, and where the binding pocket leaves the most free space.

We were also motivated by the successful application of TD to different kinds of data (see chapter 1.4.2). An especially closely related example is an evaluation of TD applied to ligand-based VS by Dieterich et al. 1994 [D94], though they had to conclude, that it was inferior to other methods.

# 3   Implementation

In order to ensure optimal performance of our implementation, we decided in favor of C++ as a programming language [H11]. The software was developed in the integrated development environment *Eclipse* with the plugin *C/C++ Development Tooling* (CDT) [W06].

Beyond some classes offered through the C++ *Standard Template Library* (STL), we employed *Eigen* for matrix, vector, and quaternion arithmetic. Eigen is an acknowledged C++ linear algebra library also used by organizations such as the *European Space Agency* (ESA) and Google [G11].

## 3.1   Software Design

For the sake of maintainability, extensibility, and flexibility, we designed the program in accordance with the object-oriented paradigm [G91]. We implemented the following classes (see also Fig. 12):

- **Atom**

  An atom has several properties specified by a line of the coordinate section of its molecule's structure file.

- **Molecule**

  Molecule is a superclass of Ligand and Receptor. A molecule has an unlimited number of atoms, can have a list of shadows and several other properties. It has a method for performing PCA on itself. The binding pocket is implemented as an instance of Molecule, since it is neither a ligand nor a receptor.

- **Ligand**

  Ligand is a subclass of Molecule. In addition to the inherited properties and methods, Ligands can be read from a structure file, have a method to compute their own density maps (in the following reffered to as shadows), and offer options of translation and rotation.

- **Receptor**

  Receptor is also a subclass of Molecule. The only additional method this class offers, is to read in a 3D protein file.

- **TangentSolver**

  This class offers methods needed for computing the TD.

- **Helper**

  The Helper class provides several methods used throughout the program. It implements various transformations of ligands, conversion of data structures, as well as writing output.

### 3.1.1 Class Diagram

The class diagram conceptually models the systematics of our program:

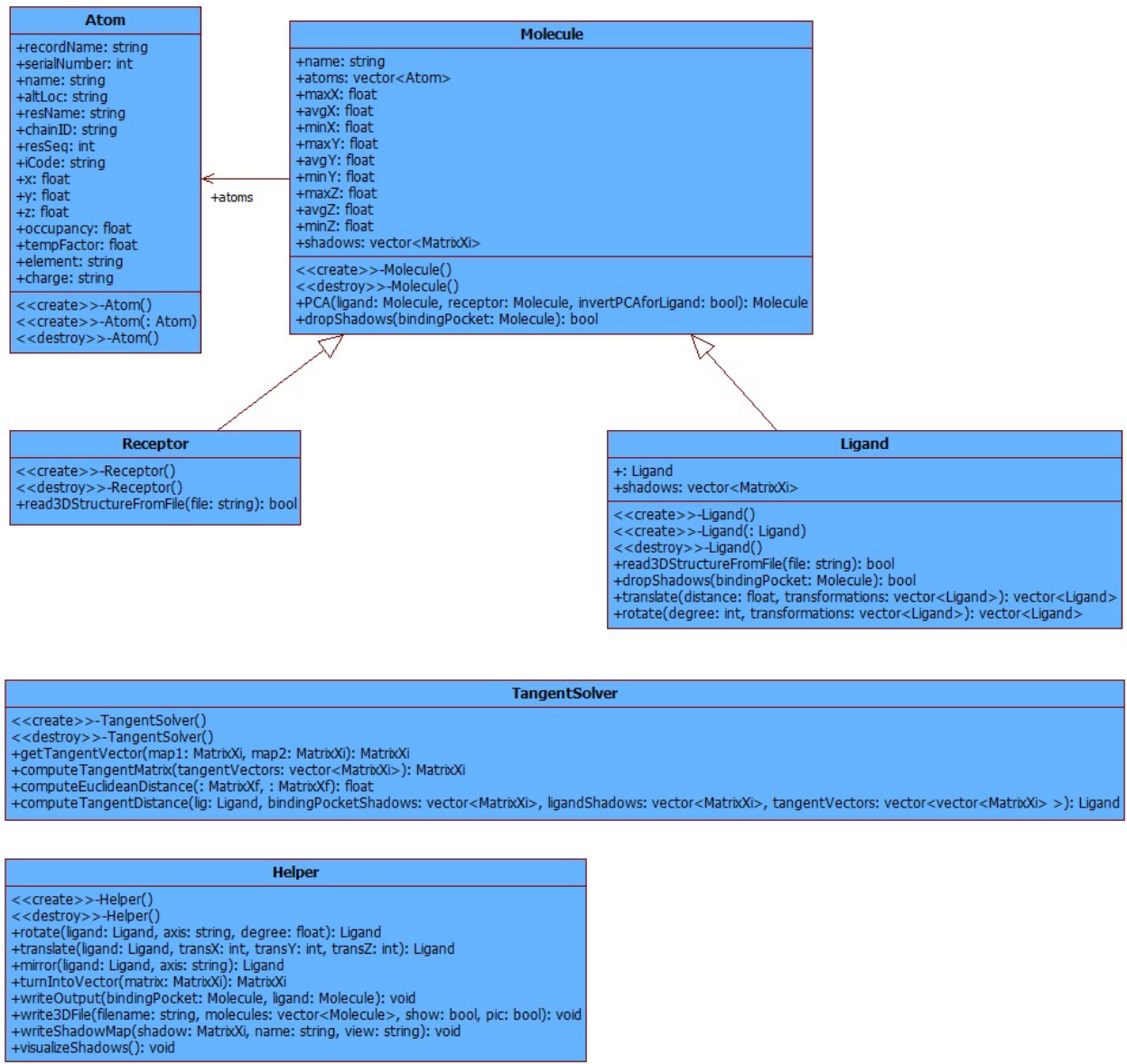

*Figure 12: Unified Modeling Language (UML) class diagram of the pre-docking filter.*

## 3.2 Program Flow

The activity diagram illustrates our program's overall flow of control:

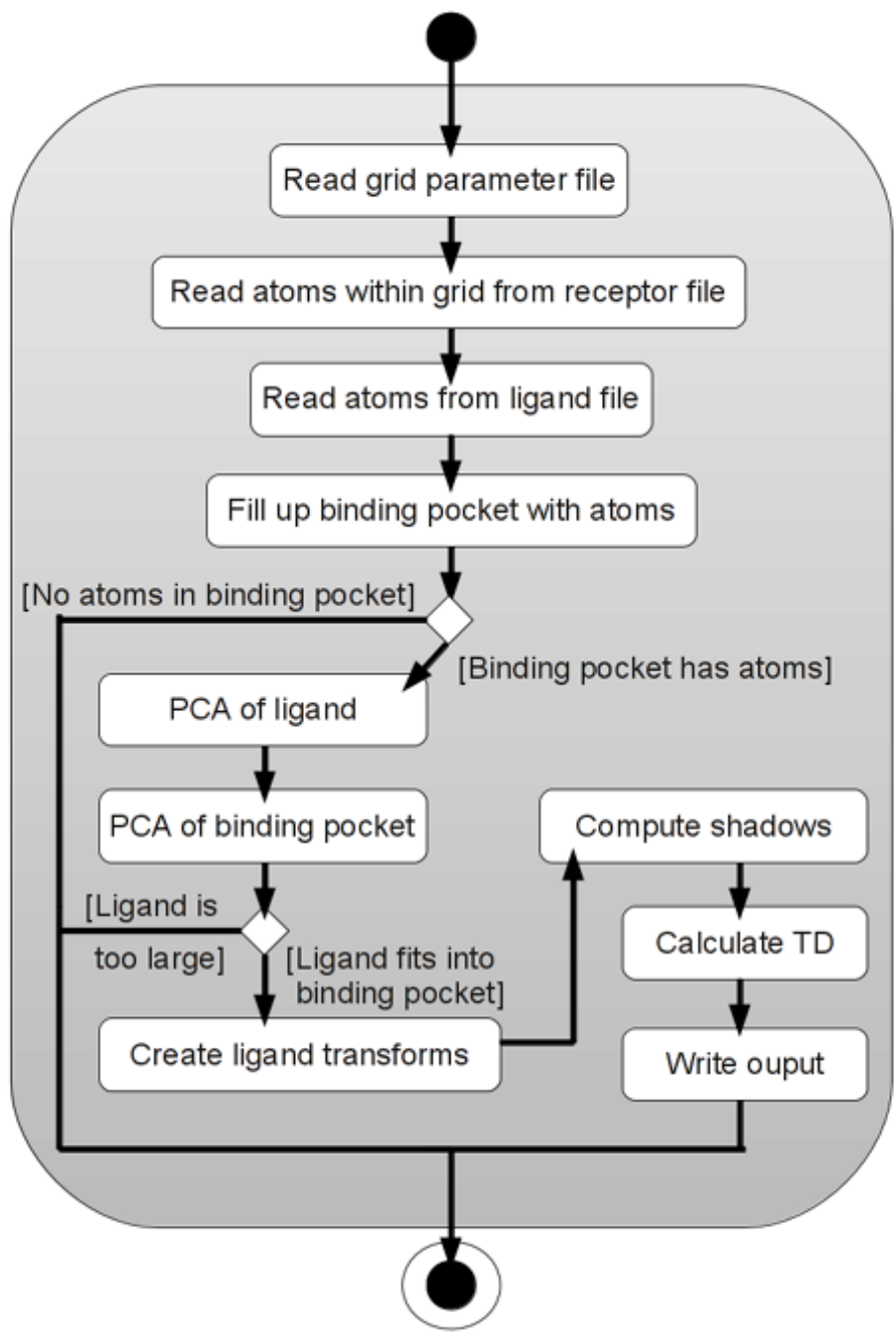

*Figure 13: UML activity diagram of the pre-docking filter.*

## 3.2.1 Input

The program expects two parameters: An identifier for the receptor and one for the ligand. If the correct number of parameters is given, it tries to open the input files following this naming scheme:

- Grid parameter file: *receptorID*.gpf
- Receptor 3D structure file: *receptorID*.pdbqt
- Ligand 3D structure file: *ligandID*.pdbqt

The grid parameter file must follow AD's specification according to [AD]. Although our program looks for coordinate files with the filename extension *pdbqt*, which stands for AutoDock 4's extension of the PDB file format, it also works with common PDB files as specified in [P11], since it does not rely on PDBQT's additional features.

When reading in the files, the relevant information is extracted: grid center, grid points, spacing, the ligand's atoms, and those atoms of the receptor located within the grid box.

## 3.2.2 Binding Pocket Identification

To generate a negative imprint of the binding pocket, the program fills it up with spheres (i.e. default atoms with a radius modified by ($0.5 \cdot$ *spacing*), to fill the grid with spheres as tightly as possible). Therefore, we visit all grid points. For each point, we iterate over all receptor atoms and compute the Euclidean distance $D_{Euclid}$ between the point and the atom. If, after subtracting the vdW radius of the atom (see Table 2), and the radius of the sphere, $D_{Euclid}$ remains $\geq 0$, then there is enough free space left to put a sphere onto the grid point. Finally, all spheres are combined in a new instance of Molecule, representing the negative imprint. For illustrations, see Fig. 14 and 15.

| Element | Radius (Å) |
|---|---|
| Carbon (C) | 1.70 |
| Hydrogen (H) | 1.20 |
| Nitrogen (N) | 1.55 |
| Oxygen (O) | 1.52 |
| Phosphorus (P) | 1.80 |
| Sulfur (S) | 1.80 |
| Default / Other | 2.00 |

*Table 2: Van der Waals radii.*
*Non-default values from Bondi 1964 [B64].*

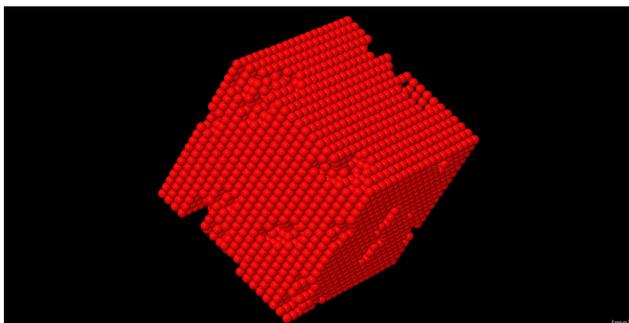

Figure 14: Visualization of spheres filling a binding box. A binding box is defined as the free space left between the atoms of a receptor within the cuboid constrained by the given grid box parameters. The grid box parameters account for the cuboid-like shape of this binding box visualization, while the atoms of the receptor within the binding box account for the small gnawed off edges and pieces. Generated with Jmol [H10].

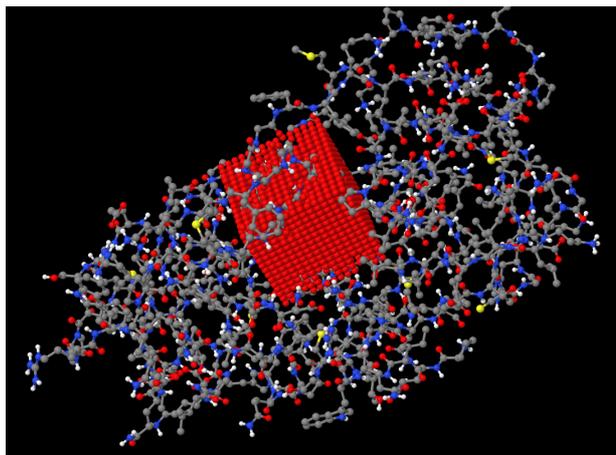

Figure 15: Superimposition of the negative imprint of the binding pocket from Fig. 14 and the receptor HIV II protease. Graphic generated with Jmol [H10].

### 3.2.3 Principal Component Analysis

A PCA is performed for both molecules to align them (pre-post comparison see Figures 16 and 17):

First, the molecules are centered around the origin by subtracting the average across each dimension from the respective dimension's individual values.

Then the data's covariance matrix $C$ is calculated according to Equations 5 and 6.

$$cov(x,y) = \frac{\Sigma_{i=1}^{n}\left(x_i - x_{avg}\right)\left(y_i - y_{avg}\right)}{(n-1)}$$

Equation 5: Calculation of the covariance between x and y in a set with n data points.

$$C = \begin{pmatrix} cov(x,x) & cov(x,y) & cov(x,z) \\ cov(y,x) & cov(y,y) & cov(y,z) \\ cov(z,x) & cov(z,y) & cov(z,z) \end{pmatrix}$$

Equation 6: Composition of the covariance matrix C.

$C$ is symmetric since $cov(x,y) = cov(y,x)$. We exploit this fact and save runtime by only calculating the diagonal values and those below, then copying those below to the upper triangle.

The next step is the computation of the eigenvalues and unit eigenvectors of *C* (see also [A07]). These are the scalars *λ* and vectors *v* fulfilling the equation *Cv = λv*. We use Eigen's class *SelfAdjointEigenSolver* for this purpose. It is an especially fast and accurate implementation of an eigendecomposition, that only works for selfadjoint (i.e. symmetric) matrices. The extracted eigenvector with the highest eigenvalue is the dominant principal component. It points in the direction of the longest stretch of the molecule.

Now, we align the molecule's longest stretch with the x-axis. Again, we make use of Eigen's advanced features, this time employing its geometry module: We generate a rotation matrix from a quaternion (for details about quaternion rotation see [H06a]), which represents a rotation sending the dominant eigenvector to the unit vector codirectional with the x-axis. The matrix is multiplied by a vector representation of the atoms' coordinates. We update the coordinates accordingly.
Since we do this with both the ligand and the binding pocket, they are aligned along their longest stretches (i.e. the x-axis) afterwards.

Finally, we repeat the whole PCA procedure, but with the purpose of additionally aligning the molecules according to their second principal components and the y-axis. The only thing we have to change about the algorithm, is to initialize the quaternion differently: this time we do it with the eigenvector with the second-highest eigenvalue and the unit vector describing the y-axis.
Since the principal components are perpendicular to each other, it is not necessary to repeat the algorithm a third time for alignment along the third-longest stretch, since there is no DOF left.

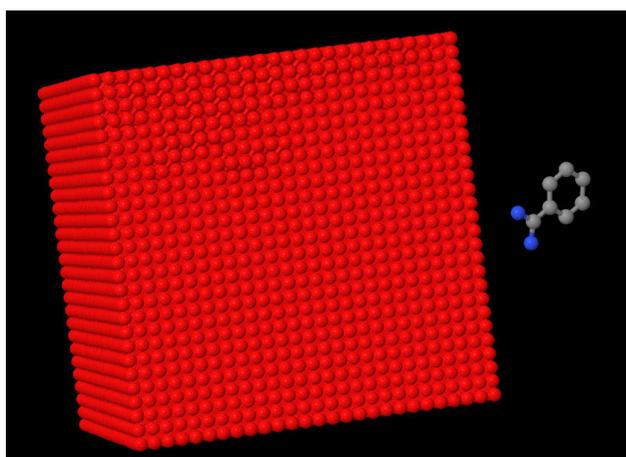

*Figure 16: Before PCA: Ligand and grid box. The ligand benzamidine (right) is located completely outside of the grid box (red spheres, left) defined for the receptor bos taurus beta-trypsin (PDB ID: 3PTB). Graphic generated with Jmol [H10].*

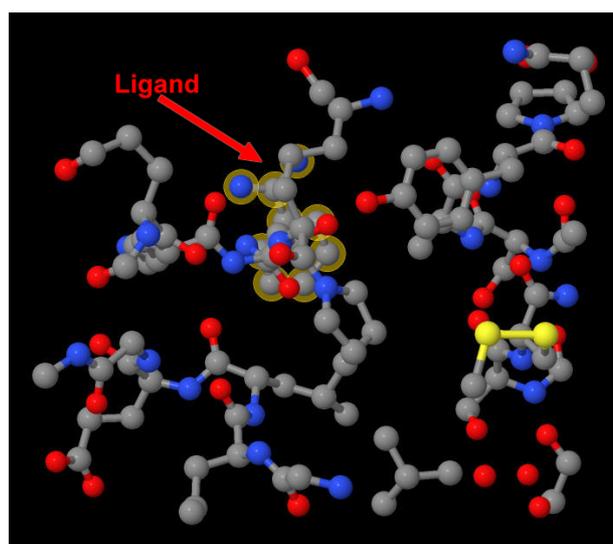

*Figure 17: After PCA: Ligand and receptor's atoms in the grid box (same perspective as Fig. 16). Now the ligand is located approx. in the middle of the grid box, illustrated by the receptor's atoms within the cuboid. Picture rendered with Jmol [H10].*

Now the time has come for filtering:

For each dimension *dim*, we compare the highest ($max_{dim}$) and lowest value ($min_{dim}$) of the ligand's coordinates with the corresponding ones of the binding pocket. If the ligand's $max_{dim}$ (/ $min_{dim}$) is higher (/ lower) than the binding pocket's $max_{dim}$ (/ $min_{dim}$), then the program stops and notifies the user, that the ligand is too large for the given binding pocket. The user has the options to continue with another ligand or try again making the grid box larger.

### 3.2.4  Tangent Distance

TD is based on TV's. In order to compute those, we first need to generate transformations of our prototypes. We decided to treat the ligand as the prototype and the binding pocket as the pattern to be classified. However, this does not make much of a difference since we do not actually use TD for classification, but only for distance minimization.

The program creates six copies of the ligand, which are then subjected to transformations:

Three are being rotated by an angle $\theta_{deg}$ = 20°, each around another axis. This is implemented by converting $\theta_{deg}$ to radians: $\theta_{rad} = \theta_{deg} \cdot 0.0174562925$ – and inserting $\theta_{rad}$ as an angle into the respective basic rotation matrix (see Equations 7-9). The other three ligand copies are translated by two grid points, each in another dimension, by adding (2 · *spacing*) to each of the respective dimension's coordinates.

$$R_x(\theta_{rad}) = \begin{pmatrix} 1 & 0 & 0 \\ 0 & \cos(\theta_{rad}) & -\sin(\theta_{rad}) \\ 0 & \sin(\theta_{rad}) & \cos(\theta_{rad}) \end{pmatrix}$$

$$R_y(\theta_{rad}) = \begin{pmatrix} \cos(\theta_{rad}) & 0 & \sin(\theta_{rad}) \\ 0 & 1 & 0 \\ -\sin(\theta_{rad}) & 0 & \cos(\theta_{rad}) \end{pmatrix}$$

$$R_z(\theta_{rad}) = \begin{pmatrix} \cos(\theta_{rad}) & -\sin(\theta_{rad}) & 0 \\ \sin(\theta_{rad}) & \cos(\theta_{rad}) & 0 \\ 0 & 0 & 1 \end{pmatrix}$$

*Equations 7-9: Matrices for 3D rotation around the x-, y-, and z-axis.*

We observed that this is a crucial step in making TD an efficient method for set-ups with manifolds: There is no need to ever combine the allowed transformations during the algorithm, nor is it necessary to perform one of the deformations several times with different parameters. These issues are solved by merging the information included in the TV's in linear combinations, and later finding the best parameters by applying basic optimization methods.

Now, we introduce the additional step that is not necessary in TD algorithms when applied to 2D data: The program generates shadows of the ligand and binding pocket from different views to account for the additional third dimension. The views considered for the binding pocket are: from front, left, above, back, right, and below. For the ligand, we just drop shadows for the first three of those views. This is based on the fact, that the views are not independent. For comparing all possible orientations, three views of one molecule and six of the other are sufficient. This can be visualized by envisaging two dice: one of them is kept still (three sides are visible), while the other is moved around it (all six sides are visible at a given time) – a valid way to produce all possible orientations. This is basically what the program does later, when minimizing the TD.

The computation of a shadow works as follows: A matrix of equal size is generated. The size is specified by the dimensions of the respective view of the grid. I.e. if the binding pocket has 45 grid points in the y-dimension and 35 in the z-dimension, then a shadow matrix for the view from left must accomodate 35 columns and 45 rows. All of the matrix's coefficients are initialized to zero.
An iteration over all atoms of the molecule increments the respective coefficient, which can be located by subtracting the value of the lower bound of the grid from the atom's value and then dividing the result by *spacing*.
Our software includes functionality to write the shadows to comma-separated values (CSV) files and visualize them as bitmaps. Some examples are given in Figures 18 and 19.

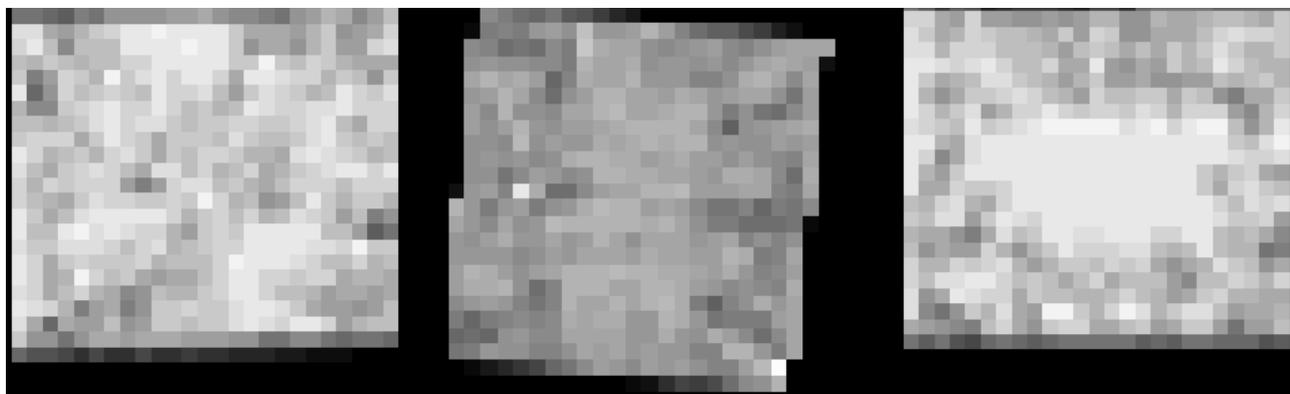

*Figure 18: Drop shadows of a binding pocket. Lighter pixels in HIV protease's binding pocket indicate more free space (and vice versa less space in its negative imprint). Left: Front. Middle: Left. Right: Above.*

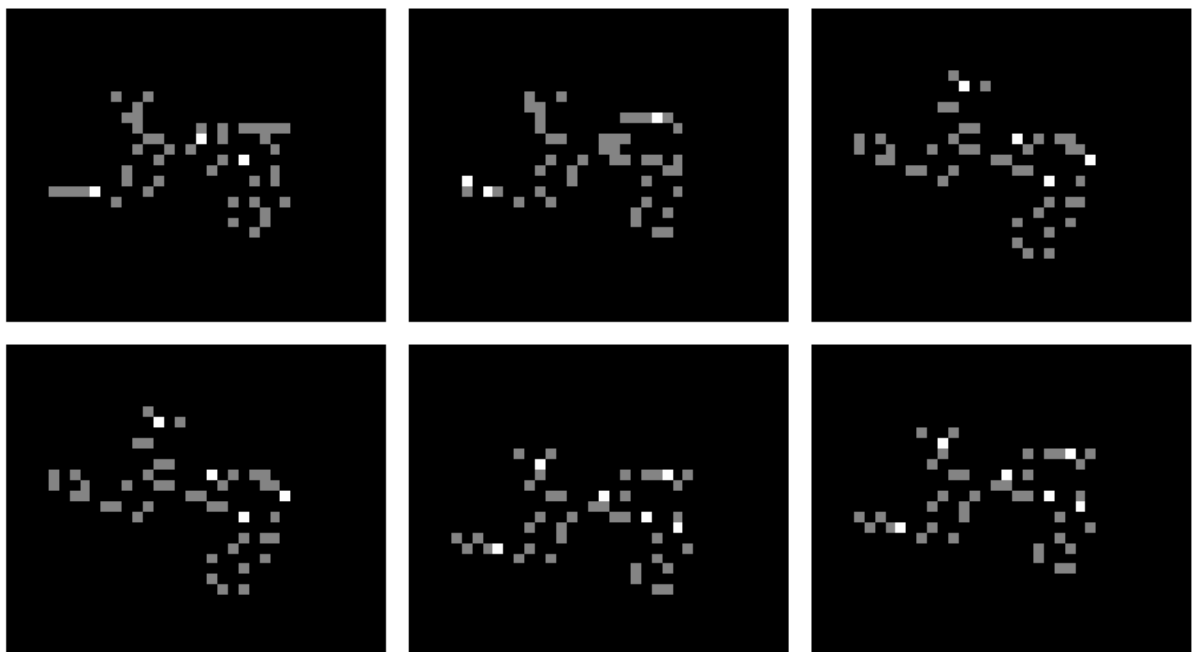

*Figure 19: Drop shadows of a ligand. Indinavir, an HIV protease inhibitor, was subjected to six transformations. These are their frontal shadows. Top left: Rotation around x-axis. Top middle: Rotation around y-axis. Top right: Rotation around z-axis. Bottom left: Translation along x-axis. Bottom middle: Translation along y-axis. Bottom right: Translation along z-axis.*

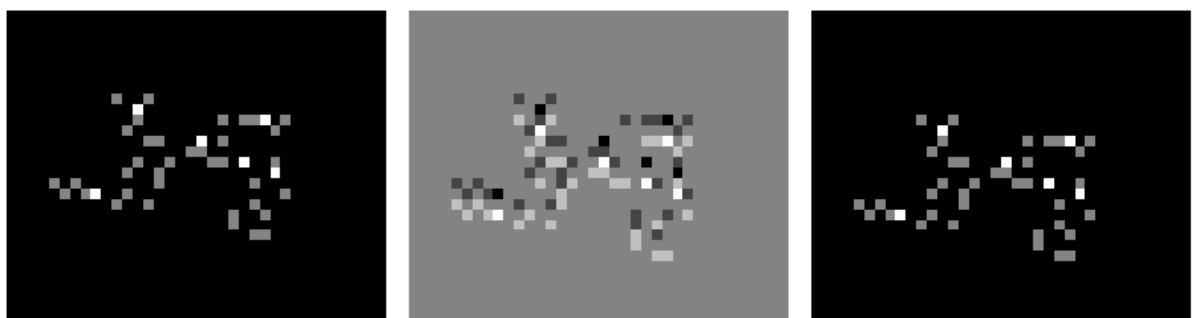

*Figure 20: TV resulting from translation. The ligand (left: original state) has been translated by 2 grid points along the y-axis (right: translated ligand). The corresponding TV is presented in the middle.*

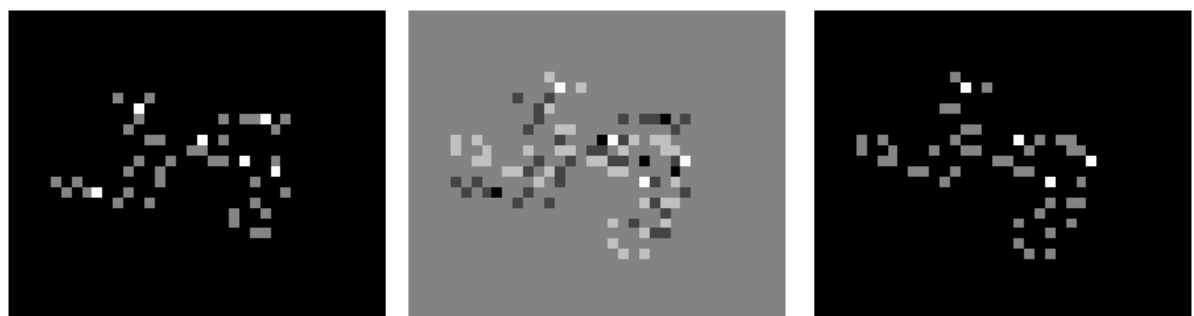

*Figure 21: TV resulting from rotation. The ligand (left: original state) has been rotated around the z-axis (right: rotated ligand). The corresponding TV is presented in the middle.*

The next step is the generation of TV's according to Equation 3. Since we have six ligand transforms and three shadows for each of them, a total of 6 · 3 = 18 TV's is computed. These vectors are also visualizable by the above-mentioned functionality (results see Figures 20 and 21).

The only two things remaining to do, before we can proceed to the TD algorithm itself, are these:

1. Generate matrix $T_i$ by putting together as columns all TV's of view $i$.
2. Create vector $a$, in which the parameters of the transformations are stored. These are the numbers we vary in order to find the minimal distance. For the rotational parameters, we try out values in the range between -90° and 90°, with steps of 15° in between. The translation parameters are extracted from the length of the respective dimension. Since these usually range somewhere between 30 and 150, we decided for an increment of 8, in order to have approx. the same resolution for translation and rotation.

Finally, we can solve Equation 4:

- Initiate all TD's to ∞.
- Iterate over the views $i$.
    - Multiply $T_i$ by $a$.
    - Compute $D_{Euclid}$ between $T_i a$ and the respective shadow of the binding pocket.
    - Compute $D_{Euclid}$ between $T_i a$ and the opposite shadow of the binding pocket (e.g. back view, when $i$ is the ligand's front view).
- Update the corresponding TD, if it is smaller in the current iteration.
- Sum up the distances of those views generating a valid description of the ligand:
    front + left + above; front + right + below; back + left + below; back + right + above
- Update the overall distance, if the lowest summed up distance is lower, and remember the corresponding views-based ligand description as well as the parameter vector $a$.

The last action here is to rotate and translate the ligand according to the result of the TD algorithm. The required information is contained within the parameter vector and the views-based description.

## 3.2.5  Output

The output of our program consists in writing a 3D structure file in PDB / PDBQT format. It contains the ligand's coordinates optimized according to the PCA-induced alignment and the TD minimization. The filename is out_*ligandID*_final.pdbqt.

# 4 Experimental Results

After having implemented the TD algorithm as a pre-docking filter, the next task at hand was to evaluate its potential. A promising program would be expected to

a) tell apart ligands that might fit into a given binding pocket from ligands that are too large,

b) provide optimized coordinates for ligands that might fit, and

c) make the overall runtime of a docking run including pre-docking filtering shorter than the runtime of a plain docking run.

Therefore our evaluation required answering the following three questions:

1. Is the filter able to distinguish ligands in terms of their potential fit?

2. How high is the quality of docking results with optimized ligand coordinates?

3. Is our implementation good enough in terms of runtime to speed up docking?

The answers we found are given in the following subchapters.

## 4.1 Evaluation of the Size-Dependent Filtering

The size-dependent filtering is performed subsequent to the alignment of the ligand and the binding pocket according to their centroidal axes. However, the filtering criterion is based on an approximation: The first principal component is a line of best fit, but it does not necessarily connect the two single most far apart atoms of a molecule – the atoms used for the filtering decision. Thus, even if the ligand's molecules do not fit into the binding pocket after the PCA-based alignment, it may well be that a slight transformation would make the ligand fit. The same applies to the second and third principal components.

This caveat demands sound testing in order to be able to assess the filter's fitness for use.

### 4.1.1 Classification Performance

We tested the size-dependent filter on a large dataset, which is part of the directory of useful decoys DUD [H06b]. DUD contains 40 packages: each includes one receptor together with an average of 70 known ligands and 36 decoys per ligand. A decoy does not bind to the given receptor. The authors applied the additional criterion, that these compounds be similar to the ligand with regard to physical properties (e.g. calculated partition coefficient, molecular weight).

We selected six packages from the DUD, resulting in a dataset of a total of 170 ligands and 6,435 decoys. Since the small molecule files were only available in mol2 format, we had to convert them to PDB format using the widespread conversion tool Open Babel [G06]. Afterwards, we applied a script written by ourselves in PHP, to split up each receptor's collective ligand / decoy file into multiple files, each of them containing only one molecule. Finally, we applied our filter to all of those, using the grid box parameters defined by the DUD authors. We scaled the boxes' sidelengths down by 25% each, which yielded better classification results (i.e. not only negatives).

A filter is supposed to reject a higher rate of decoys (true negatives / TN) than ligands (false negatives / FN), and vice versa let pass a lower rate of decoys (false positives / FP) than ligands (true positives / TP). The individual results are listed in Table 3. Figures 22 and 23 make the filtering performance visually comparable.

| Receptor | Ligands | | | | Decoys | | | |
|---|---|---|---|---|---|---|---|---|
| | Count | | Percent | | Count | | Percent | |
| | TP | FN | TP | FN | TN | FP | TN | FP |
| Angiotensin-converting enzyme (ACE) | 46 | 1 | 97.87% | 2.13% | 68 | 1,657 | 3.94% | 96.06% |
| Glucocorticoid receptor (GR) | 56 | 20 | 73.68% | 26.32% | 775 | 2,020 | 27.73% | 72.27% |
| HIV reverse transcriptase (HIV RT) | 37 | 1 | 97.37% | 2.63% | 3 | 1,432 | 0.21% | 99.79% |
| Mineralcorticoid receptor (MR) | 12 | 1 | 92.31% | 7.69% | 25 | 508 | 4.69% | 95.31% |
| Progesterone receptor (PR) | 22 | 3 | 88.00% | 12.00% | 352 | 613 | 36.48% | 63.52% |
| Retinoic X receptor alpha (RXR) | 11 | 7 | 61.11% | 38.89% | 393 | 314 | 55.59% | 44.41% |
| Totals | 138 | 32 | 82.49% | 17.51% | 1548 | 4,887 | 24.94% | 75.06% |
| | 170 | | 100.00% | | 6,435 | | 100.00% | |

*Table 3: Results of size-filtering the selected molecules from the DUD set.*

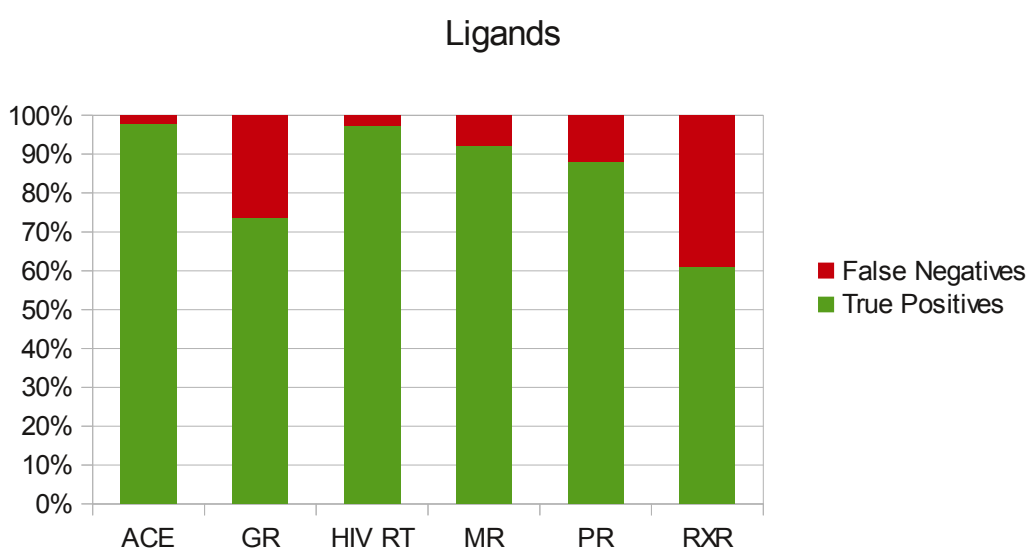

*Figure 22: FN and TP after size-filtering the DUD ligands.*

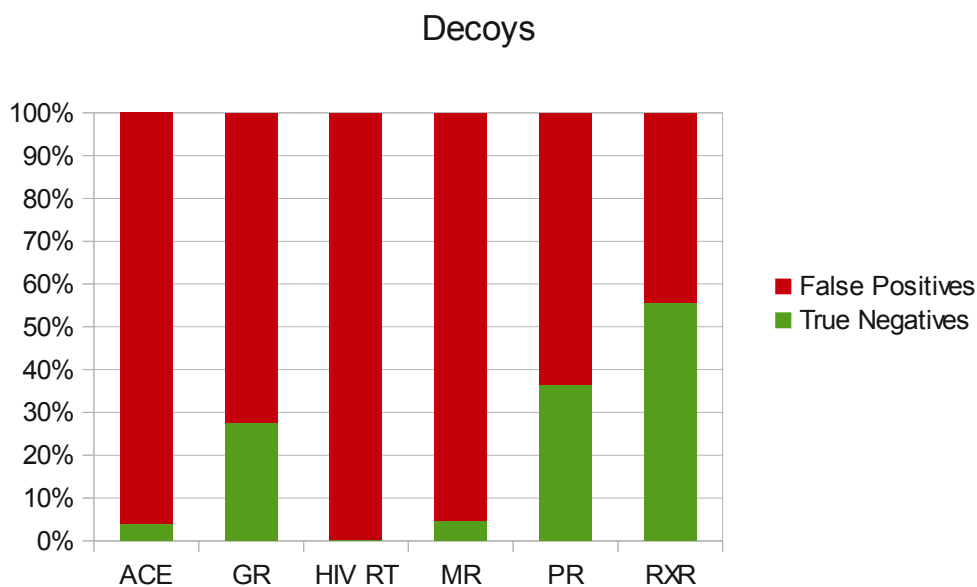

*Figure 23: FP and TN after size-filtering the DUD decoys.*

We analyzed the raw data with regard to statistical measures of our classifier's performance: specificity, sensitivity, and accuracy. These are the results:

$$Specificity \; = \; \frac{\Sigma\,TN}{\Sigma\,TN \; + \; \Sigma FP} \; = \; 0.24$$

$$Sensitivity \; = \; \frac{\Sigma\,TP}{\Sigma\,TP \; + \; \Sigma FN} \; = \; 0.81$$

$$Accuracy \; = \; \frac{\Sigma\,TP \; + \; \Sigma\,TN}{\Sigma\,TP \; + \; \Sigma FP \; + \; \Sigma FN \; + \; \Sigma\,TN} \; = \; 0.26$$

*Equations 10-12: Equations and results for specificity, sensitivity, and accuracy.*

## 4.1.2   Quality Evaluation of the Start Orientation Optimization

Beyond computing classification performance values for the filtering criterion itself, we examined whether the PCA-inherent alignment of ligand and binding pocket also produced superior start conformations for subsequent docking runs, even before applying the TD algorithm. We will describe this validation process in the following.

The fitness of a docking result is expressed by its docking score. Different docking software usually implements different scoring functions. In order not to run into problems when comparing docking results, we decided to use one software throughout all of our docking-related evaluation experiments: AutoDock 4.2.3.

To analyze the output of the size filter we developed, we relied on a tried and true (see also Morris et al. 2009 [M09b], Niv & Weinstein 2005 [N05]) method for the validation of docking results: the so-called redocking. This procedure is based on the idea to dock a known ligand back to the native structure of the corresponding receptor when in a complex.

This is a step-by-step description of redocking:

1. Find a PDB file containing the receptor with the ligand co-crystallized in its binding site.
2. Split the file in two files by cutting out the lines describing the ligand's coordinates and inserting them into a separate file.
3. Randomly change the ligand's orientation.
4. Perform a docking run with the two files resulting from step 2 and 3.

There are two major differences between redocking and regular docking:

1. When extracting the receptor structure from a file where the protein is complexed with the ligand, the protein's coordinates can be assumed to be in accordance with the *induced fit* – the fit induced by small changes in conformation resulting from interactions between the receptor and ligand during binding. This is an advantage, since this fit presumably better reflects the biochemical reality, than a 3D structure resulting from the receptor alone.
2. On the other hand we have the comparability of the results: A validation of the redocked pose can easily be achieved by analyzing the root-mean-square deviation (RMSD) between this ligand's atomic coordinates and those of the original docked ligand.

However, these advantages do not apply to the common use cases of docking, where the procedure has to be performed without prior knowledge about the correct fit. Therefore the exploitability of the strengths of redocking are limited to these scenarios: the verification of docking results, where the directive is being able to judge the performance of the procedure compared to others – and alternatively the validation of chosen docking parameters. In the latter case, redocking can help decide whether to take which trade-off regarding runtime versus accuracy of results.

### 4.1.2.1 Ligand Preparation

First, the ligands were randomly translated and rotated. This is necessary in redocking since the original atomic coordinates of each ligand come completely optimized, while the purpose of the procedure however is testing the optimization performance of a given tool.

The further preparation procedure followed the AD preprocessing routine using ADT 1.5.4:

1. We added all hydrogen atoms to the molecule in order to satisfy all the valences of the heavy atoms.
2. We let ADT compute partial charges applying the method developed by Gasteiger & Marsili [G80].
3. We deleted the non-polar hydrogen atoms and merged their charges with the carbon atoms in order to satisfy the united-atom model according to Weiner et al. 1984 [W84] on which AD relies for representing molecules.
4. We used the ADT feature to automatically assign atom types, which for the special case of AutoDock 4 include information about an atom being a hydrogen bond acceptor or donor, as well as being aromatic or aliphatic in the case of a carbon atom. If an atom belongs to none of these categories, the atom type simply is the same as its element.
5. We saved the ligand to a file in PDBQT format.

We did not use ADT's option to define rotatable bonds and the AD-specific torsion tree representation of the potential torsions of the ligand, because when performing redocking, it is not necessary to account for ligand flexibility. Simply treating the ligand extracted from the complex as a rigid structure, achieves best comparability with the native complex, since the ligand will have the same conformation both in the complex and in the docked poses.

## 4.1.2.2  Receptor Preparation

The receptors also needed some preparation:

1. We added missing hydrogen atoms.
2. We computed Gasteiger charges.
3. We merged non-polar hydrogen atoms and charges.
4. We assigned atom types.
5. We saved the receptor to a file in PDBQT format.

Although AD offers flexible treatment of individual amino acids, we used the common method of treating the receptor as a rigid structure. The reasons are basically the same as the above-mentioned reasons for using a rigid ligand.

## 4.1.2.3 Redocking

We selected two ligand / receptor pairs for redocking:

1. **Flaviviral protease** (PDB-ID: 2FOM) / **glycerol** (ID: GOL)

   **Grid parameters:** Spacing = 0.375. Extension: 40x40x40. Center: 2.549, 7.527, 23.767.

2. **Trypsin** (PDB-ID: 3PTB) / **benzamidine** (ID: BEN)

   **Grid parameters:** Spacing = 0.375. Extension: 50x50x50. Center: -4.451, -10.488, 17.074.

In order to approximate our program's baundary conditions with these tests, we created a very tightly fitting grid box for benzamidine, and a larger grid box for the even smaller ligand glycerol. This is useful since the size of the grid box is the critical parameter with regard to the runtime of our filter: Its complexity is $O(n \cdot m)$, where $n$ depends on the number of grid points, and $m$ on the number of the receptor's atoms located within the binding box.

We used the same docking protocol for both settings: Lamarckian genetic algorithm, number of runs: 1, all other settings: ADT default values.

We measured the runtime our filter needed to preprocess the data, without the TD algorithm. Then we took measurements of AD runs starting with differently oriented ligands:

1. Randomized orientation
2. PCA-optimized orientation

We repeated each procedure three times and averaged the runtimes for more reliable results.

Another measure we were interested in, was the RMSD of the ligand's coordinates when co-crystallized, and those resulting from our optimization. Results are listed in Table 4.

| Receptor / Ligand | Runtime (seconds) | | | | Runtime (%) | | RMSD |
|---|---|---|---|---|---|---|---|
| | Our tool | AD | Total | AD only | Total vs. AD only | AD vs. AD only | |
| 2FOM / GOL | 4.0690 | 18.3790 | 22.4480 | 19.3194 | 116.19% | 95,13% | 31.99 |
| 3PTB / BEN | 1.2387 | 15.1920 | 16.4307 | 15.1445 | 108.49% | 100.31% | 10.09 |

*Table 4: Evaluation of the performance of the PCA-based optimization of the ligand orientation. RMSD in Å.*

## 4.2 Evaluation of the Tangent Distance Optimization

The TD algorithm was tested with a set of receptors, ligands, and grid boxes based on a selection made by Wilantho et al. 2008 [W08]. They also developed a shape-matching-based pre-docking filter. Their collection was chosen with the intent to have exceptionally high variance in the receptors' binding sites' shapes and sizes. This variety made sense in our test case of a shape-matching algorithm, too.

We chose a subset of eleven receptor / ligand pairs (identifiers itemized in [W08]) and added the already introduced combinations of HIV / indinavir (MK1) and trypsin (3PTB) / benzamidine (BEN) to it. In order to start our tests from an especially low-level task, we performed redocking, without prior randomization of the ligand pose. I.e. we ran our program on the ligand and receptor with their co-crystallized orientation. It can be expected that a well-functioning optimization procedure solves this task best compared to more complicated settings. Afterwards, we computed the RMSD between the original ligand's coordinates and those resulting from the TD calculations.

The achieved results are presented in Table 5, as well as in Figures 24 and 25.

| Receptor | Ligand | Binding site | | TD | RMSD |
| | | Size | Shape | | |
|---|---|---|---|---|---|
| 1R6A | RVP | Small | Complex | 8,902,009 | 8.841 |
| 1HSG | MK1 | Large | Simple | 30,000,000 | 10.451 |
| 3PTB | BEN | Small | Simple | 6,626,073 | 12.302 |
| 1CET | CLQ | Small | Simple | 6,539,446 | 15.507 |
| 1ZNY | GDP | Small | Simple | 6,623,228 | 16.281 |
| 2C27 | MA8 | Large | L-shape | 6,579,995 | 16.334 |
| 1EYE | PMM | Large | Simple | 6,513,225 | 18.516 |
| 1RQ7 | GDP | Small | Complex | 6,551,944 | 18.523 |
| 1ZAU | AMP | Small | Simple | 6,542,517 | 19.651 |
| 1MRN | T5A | Large | L-shape | 15,359,524 | 19.677 |
| 1ENY | NAD | Large | Complex | 11,302,646 | 19.711 |
| 1DF7 | MTX | Small | L-shape | 6,535,350 | 20.193 |
| 2FOM | GOL | Small | Simple | 15,579,435 | 24.713 |

*Table 5: Results of the TD-based ligand orientation optimization. The list is ordered by RMSD (in Å), ascending.*

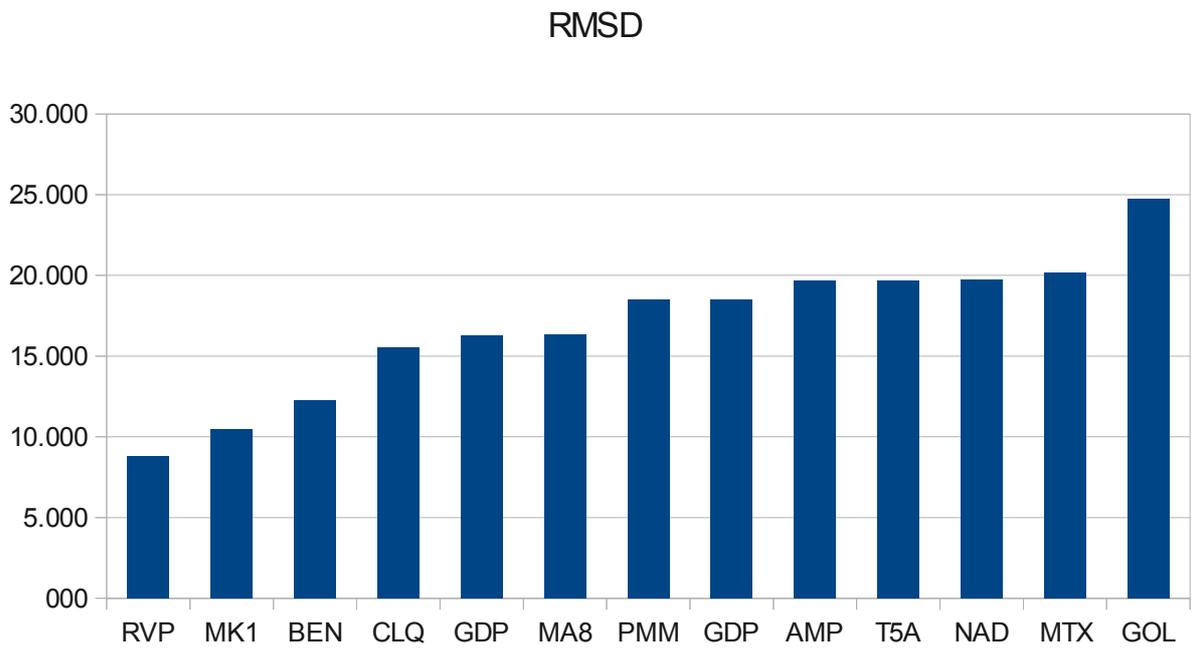

*Figure 24: RMSD values of TD optimization in redocking. RMSD in Å.*

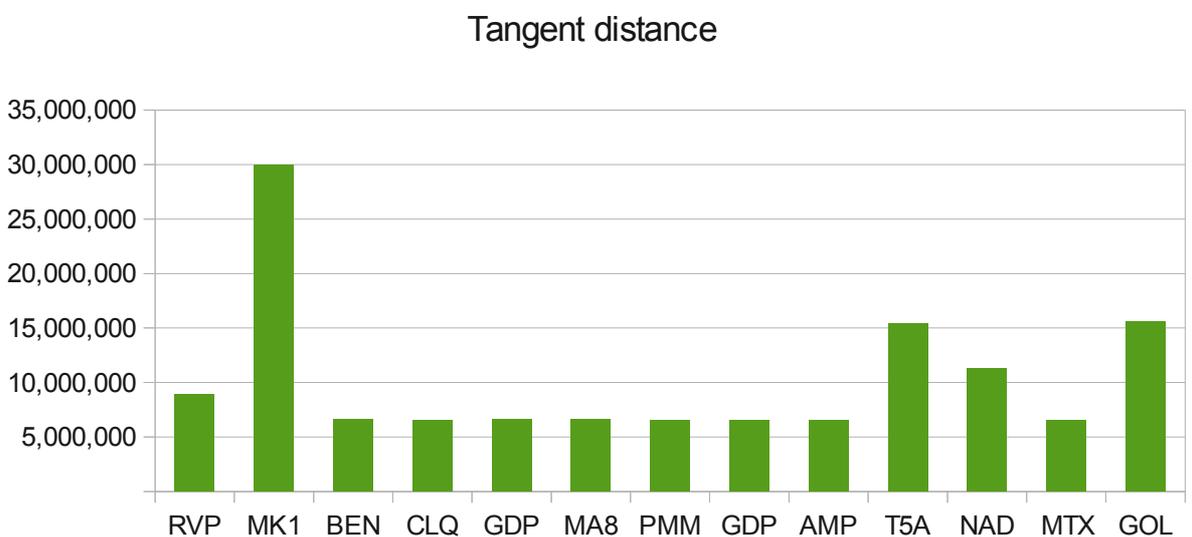

*Figure 25: Tangent distances resulting from TD optimization. The columns are ordered by ascending RMSD.*

Since no correlation was found between the exceptionally high RMSD values and any of the other properties monitored, and since we discovered a very plausible explanation for these results, no further tests were performed with the TD part of our program.

# 5   Conclusions and Outlook

For the discussion of the experiments, we revisit the questions asked in chapter 4, and present our answers based on the experimental results:

1.  **Is the filter able to distinguish ligands in terms of their potential fit?**

    The filtering part of our implementation was evaluated on a test set taken from the DUD. The classification performance showed a high degree of variance between different packages: For RXR, for example, 55.59% of the decoys were classified correctly, while only 0.21% of the HIV RT decoys were rejected by the filter. However, whenever the rate of TN was very high, the rate of TP was also very high, leading to a balance in the overall accuracy. Therefore we averaged the specificity, sensitivity, and accuracy over all results.

    Usually, a trade-off has to made between high specificity and high sensitivity, depending on whether FN or FP would be regarded a more critical error. In our case, we have compounds for VS available in abundance, but only very few of those turn out to be ligands. This leads to the conclusion, that in the given scenario, it is much more important not to reject ligands (FN), than not to include some decoys (FP). The measure that focusses on the recognition rate of TP vs. FP is sensitivity. While our classifier's specificity of 0.42 and accuracy of  0.26 are not very high, its sensitivity of 0.82 can be interpreted as a sign, that the filter's performance has a certain potential.

    Another positive result is that our filter achieved an enrichment in terms of the percentage of ligands in the datasets. It was increased from 2.64% in the original DUD data to 2.82% in the set of compounds that passed the filter.

2.  **How high is the quality of docking results with optimized ligand coordinates?**

    Answering this question requires a separate examination of the components of our tool: First, there is the optimization based on PCA – second, we try to further optimize the coordinates with the TD measure.

The optimization with PCA was tested by redocking and RMSD evaluation of two receptor / ligand pairs. The higher RMSD of 31.99Å resulted from a test with a large grid box, while the lower (but still not exactly promising) RMSD of 10.09Å was linked to a run with a smaller box. Based on how the PCA-induced coordinate optimization works, and on the above-mentioned results, we do not state a high quality in this approach's optimization, because the results largely depend on the choice of grid box parameters. However, an approximate overlap of the binding pocket and ligand was achieved (see also Figures 16 and 17) as a byproduct of the method that was only intended as a filtering criterion.

In the evaluation of the TD-based shape-matching, we found RMSD values in a range between 8.841Å and 24.713Å. A promising method for pose optimization can be expected to yield an average RMSD of about 5.0Å or less. This would have been a threshold, where it would have made sense to try to improve the algorithm in terms of runtime and accuracy. However, since TD only produced orientations with quite high RMSD values, and there was not even a correlation between RMSD and any of the other observed properties like binding site size, binding site shape, and TD itself, it seems plausible that the algorithm may just not be suitable for the given problem.

Thinking about and investigating possible causes, we found two probable reasons explaining the failure of TD in vague binding pocket / ligand shape-matching:

1.  TD is usually applied to 2D data, mostly images. It is known to perform well in these settings. Since we wanted to solve a problem in 3-space using TD, we came up with the idea to use a set of density maps as a 2D representation of our 3D data. The handling of the emerging multitude of views and shadows of different sizes, especially when it comes to comparing them to each other choosing only reasonable combinations, is difficult. Furthermore, the chosen representation may lack preciseness, since the shadows disregard one of three dimensions. The combination of shadows from different views may not make up for this loss.

2.  The more important assumed weak point in the approach is directly connected to a known limitation of TD: Several authors report that TD only yields good results, if the transformations that it has to deal with are reasonably small (see e.g. [V05]). This is without much doubt not the case when comparing a ligand to the negative imprint of a binding pocket. Although the ligand fits into the shape described by the imprint, both usually have little in common regarding their overall shape.

**3.  Is our implementation good enough in terms of runtime to speed up docking?**

Since the evaluation of the TD approach revealed its inapplicability to pre-docking optimization, we did not further investigate the runtime of this component of our software.

However, the PCA part succeeded in enriching the screened dataset with regard to ligands, so we tested this component's computational performance separately. Depending on the volume of the binding pocket, our program needed between approx. 1-4 seconds for processing a ligand / receptor combination. The combined runtime of a filter run and subsequent AD run took 108-116% of the time of a stand-alone run of AD. Comparing only the subsequent AD run with the stand-alone AD run evaluated to 95-100%, i.e. the AD run following our pre-docking filtering was faster on average.

Transferring the TD algorithm not only from one area of research (OCR) to a different one (molecular docking), but also from 2D data to 3D data, was an innovative and speculative idea. The aim of this thesis was to implement and evaluate not only this idea, but additionally a PCA-based filtering of potential ligands.

TD did not produce meaningful results, which is assumed to be due to the limited accuracy of the approach when applied to compare data with high variance, such as binding pockets versus ligands.

The PCA algorithm, well-known as a statistical method, succeeded in its binary classification task to a certain extent: In a virtual screening, an enrichment regarding ligands was achieved. Also, the runtime of this method is reasonably short, compared to the overall runtime required for a docking run using the most efficient algorithm currently available in AD. Despite the relatively high complexity of $O(n^2)$ of the binding pocket identification, the time this step takes could be reduced significantly by parallelizing the workload using multithreading. Thus resource utilization on modern multicore central processing units (CPU's) could be optimized.

# Table of Abbreviations

| | |
|---|---|
| **2D** | two-dimensional |
| **3D** | three-dimensional |
| **ACE** | angiotensin-converting enzyme |
| **AD** | AutoDock |
| **ADT** | AutoDockTools |
| **CDT** | C/C++ Development Tooling |
| **CPU** | central processing unit |
| **CSV** | comma-separated values |
| **DOF** | degrees of freedom |
| **DUD** | (a) directory of useful decoys |
| **FN** | false negative(s) |
| **FP** | dfalse positive(s) |
| **GR** | glucocorticoid receptor |
| **HIV** | human immunodeficiency virus |
| **MD** | molecular dynamics |
| **MR** | mineralcorticoid receptor |
| **PCA** | principal component analysis |
| **PDB** | Protein Data Bank |
| **PDBQT** | AutoDock 4's 3D coordinate format |
| **PR** | progesterone receptor |
| **QSAR** | quantitative structure-activity relationship |
| **RMSD** | root-mean-square deviation |
| **RT** | reverse transcriptase |
| **RXR** | retinoic X receptor alpha |
| **STL** | Standard Template Library |
| **TD** | tangent distance |
| **TN** | true negative(s) |
| **TP** | true positive(s) |
| **TV** | tangent vector |
| **UML** | Unified Modeling Language |
| **USPS** | United States Postal Service |
| **vdW** | van der Waals |
| **VS** | virtual screening |

# Glossary

**Binding mode**  The 3D orientation and conformation of molecules which form a complex.

**Combinatorial docking**  Structure prediction of multi-molecular assemblies, also known as macromolecular docking.

**Decoy**  A compound used in the evaluation of docking methods. A decoy does not bind to a given receptor. The DUD decoys have physical properties similar to those of known ligands.

**Docking**  A prediction method for the orientation of one molecule to a second when bound to each other to form a stable complex.

**Druglikeness**  Properties of a compound involving chemical and physical features, that increase its potential as a drug.

**Euclidean distance**  A metric for the distance between two points or vectors.

**Induced fit**  A model for the formation of a protein-ligand complex. It assumes that ligand and complex both undergo conformational changes during the process of molecular recognition due to interactions.

**Ligand**  A smaller molecule which binds to a receptor to form a complex.

**Linear combination**  A vector that can be expressed by applying vector addition and scalar multiplication to a given set of other vectors.

**MD simulation**  A computer-based physics imitation of atomic movements and interactions.

**PDBQT format**  AutoDock 4's 3D molecule structure file format. It is very similar to the PDB file format and additionally contains entries for each atom's partial charge ($Q$) and AutoDock4 atom type ($T$).

**Pose**  A potential binding mode.

**Quaternions**    Extension of the complex numbers also used for calculating 3D rotations.

**Ranking**    Creating a list of poses resulting from a docking run ordered by their scores.

**Receptor**    The *receiving* molecule in the formation of a complex with a smaller ligand; In most cases a protein.

**Redocking**    A validation method for docking methods using ligand coordinates derived from a complex of the receptor co-crystallized with the ligand.

**Scoring**    Calculation of a pose's free energy of binding according to a scoring function

**Tangent distance**    A distance measure that is applicable to high-dimensional data.

**Tangent vector**    A vector that is the tangent to a curve or plane.

**vdW radius**    Radius of the spherical-shaped space occupied by an atom

**Virtual screening**    Computer-based method to scan compound libraries for potential drugs.